\documentclass[journal]{IEEEtran}
\IEEEoverridecommandlockouts
\usepackage{float}
\usepackage{setspace}
\usepackage{graphicx}
\usepackage{subfigure}
\usepackage{cite}
\usepackage{lipsum}
\usepackage{cuted}
\usepackage{amsmath,amssymb,amsfonts}\allowdisplaybreaks
\usepackage{mathrsfs}
\usepackage{amsthm}
\usepackage{algorithm}
\usepackage{algorithmic}
\graphicspath{{./figures/}}
\usepackage{textcomp}
\usepackage{xcolor}
\usepackage{verbatim}
\usepackage{dsfont}
\usepackage{soul}
\usepackage{enumerate}
\usepackage{diagbox}
\usepackage{bm}
\usepackage{multicol}
\usepackage{booktabs}
\usepackage{tabularx}
\usepackage{ragged2e}

\newcommand{\cB}{\mathcal{B}}\newcommand{\cD}{\mathcal{D}}

\newcommand{\cO}{\mathcal{O}}
\newcommand{\cS}{\mathcal{S}}\newcommand{\cT}{\mathcal{T}}
\newcommand{\cU}{\mathcal{U}}\newcommand{\cX}{\mathcal{X}}
\newcommand{\cY}{\mathcal{Y}}

\newtheorem{myth}{Theorem}

\begin{document}

\title{Energy-Efficient Federated Learning and Migration in Digital Twin Edge Networks}
\author{
Yuzhi Zhou,~\IEEEmembership{Student Member,~IEEE},
~Yaru Fu,~\IEEEmembership{Member,~IEEE},
~Zheng Shi,~\IEEEmembership{Member,~IEEE}\\
Howard H. Yang,~\IEEEmembership{Senior Member,~IEEE},
~Kevin Hung,~\IEEEmembership{Senior Member,~IEEE},
~Yan Zhang,~\IEEEmembership{Fellow,~IEEE}

\thanks{
Y. Zhou, Y. Fu, and K. Hung are with the School of Science and Technology, Hong Kong Metropolitan University, Hong Kong, 999077, China (e-mail: s1315400@live.hkmu.edu.hk; yfu@hkmu.edu.hk; khung@hkmu.edu.hk).

Z. Shi is with the School of Intelligent Systems Science and Engineering, Jinan University, Zhuhai 519070, China (e-mail: zhengshi@jnu.edu.cn).

H. H. Yang is with the Zhejiang University/University of Illinois Urbana Champaign Institute, Zhejiang University, Haining 314400, China, and also with the National Mobile Communications Research Laboratory, Southeast University, Nanjing 211111, China (email: haoyang@intl.zju.edu.cn).

Y. Zhang is with the Department of Informatics, University of Oslo (e-mail: yanzhang@ieee.org).
}
}

\markboth{}%
{Shell \MakeLowercase{\textit{et al.}}: Bare Demo of IEEEtran.cls for IEEE Journals}
\maketitle

\begin{abstract}
The digital twin edge network (DITEN) is a significant paradigm in the sixth-generation wireless system (6G) that aims to organize well-developed infrastructures to meet the requirements of evolving application scenarios. However, the impact of the interaction between the long-term DITEN maintenance and detailed digital twin tasks, which often entail privacy considerations, is commonly overlooked in current research. This paper addresses this issue by introducing a problem of digital twin association and historical data allocation for a federated learning (FL) task within DITEN.
To achieve this goal, we start by introducing a closed-form function to predict the training accuracy of the FL task, referring to it as the data utility. Subsequently, we carry out comprehensive convergence analyses on the proposed FL methodology.
Our objective is to jointly optimize the data utility of the digital twin-empowered FL task and the energy costs incurred by the long-term DITEN maintenance, encompassing FL model training, data synchronization, and twin migration. To tackle the aforementioned challenge, we present an optimization-driven learning algorithm that effectively identifies optimized solutions for the formulated problem. Numerical results demonstrate that our proposed algorithm outperforms various baseline approaches.
\end{abstract}


\section{Introduction}
\label{sec1}

In the envisioned sixth-generation wireless systems (6G), diverse, stringent, and even conflicting communication and computation requirements are expected for connected intelligent devices \cite{DT_envision_1,DT_self1}.
In response, researchers have proposed the concept of digital twin edge networks (DITENs) to address these challenges \cite{DT_envision_2,DT_self2}.
DITEN is a cyber-physical paradigm that aims to profile intelligent physical devices in the digital plane. Through continuous, bidirectional, and automatic interaction with their physical counterparts, digital twins utilize device data, maintain up-to-date models, and predict device actions to optimize decision-making processes \cite{DT_envision_3,DT_envision_4,DT_self3,DT_self4}. Ultimately, the system's resources can be efficiently and effectively orchestrated to fulfill the requirements of future 6G networks.

However, the paradigm mentioned above requires numerous data exchanges between the physical and digital spaces. Due to the computing and storage limitations of intelligent physical devices, particularly mobile ones, digital twins are commonly established separately from their corresponding physical devices.
Consequently, data exchanges occur from users and their devices to the infrastructures of 6G networks, raising privacy concerns.
To address this issue,
many researchers have attempted to utilize federated learning (FL) while addressing problems in the design of DITENs to enhance privacy and security.
For instance, FL can be utilized to ensure user privacy in the user selection/participation problem of DITENs. Specifically,
in \cite{DT_user_selection_2}, the authors aimed to balance the learning accuracy and learning cost FL incurs in DITENs. They focused on optimizing user participation in FL tasks to optimize this trade-off. A deep reinforcement learning (DRL)-driven decision-making algorithm was developed to solve this problem.
Furthermore, in \cite{DT_user_selection_1}, FL was leveraged to establish DITENs. The authors optimized network resource allocation and user selection to minimize the variances between the average latency and the latency experienced by each user, thereby enhancing system efficiency. They also utilized a deep neural network (DNN)-based exploitation method to assist in the optimization procedure.
Moreover, by optimizing network resources and user selections, the global model broadcasting delay of digital twin-assisted FL was investigated in \cite{DT_user_selection_3}. Therein, the authors used a deep Q-learning network to solve this problem.
In terms of loss function minimization, in \cite{DT_user_selection_4}, a digital twin-empowered FL procedure with multiple aggregation stages was studied. The optimization of network resource allocation and user selection aimed to minimize the communication cost and the loss function of FL. To tackle this problem, the authors proposed a multiple-stage algorithm incorporating DRL, group swap matching, and Lagrange optimization.

On the other hand, the FL configurations, such as the selection of aggregators, resource allocations, and optimizations of aggregation frequency, were also investigated for the DITENs.
For example, the authors in \cite{DT_traditional_3} minimized the latency and the energy costs associated with the DITEN facilitating multi-tier FL, where local training and aggregation were allocated to heterogeneous edge or cloud devices. They adopted a DRL-based algorithm with offline training to achieve these goals.
Meanwhile, the authors in \cite{DT_trust_aggregate} proposed a deviated FL formulation between digital twins and users of the DITEN. To bypass this deviation, they introduced a trust-based aggregation scheme. A DRL-based aggregation frequency selection method was proposed to minimize the loss function of the FL procedure.
Apart from these, some researchers have also integrated blockchain into the FL procedure to further enhance the reliability of the DITEN \cite{DT_blockchain_1,DT_blockchain_4,DT_blockchain_5,DT_blockchain_7,DT_blockchain_8}.
However, unlike conventional FL tasks, the DITEN operates as a long-term system.
Specifically, it continuously processes newly generated data and updated models. Jointly considering the mobility of users, this procedure can result in the migration of digital twins among multiple edge servers, thereby introducing additional complexities. Unfortunately,
due to the lack of continuity with the aspect of digital twin tasks,
this concern was overlooked in the above researches \cite{DT_user_selection_1, DT_user_selection_3, DT_user_selection_4, DT_trust_aggregate, DT_user_selection_2, DT_traditional_3, DT_blockchain_1, DT_blockchain_4, DT_blockchain_5,DT_blockchain_7,DT_blockchain_8}.
For instance, \cite{DT_blockchain_7} proposed a DRL-based algorithm to solve the resource allocation and energy harvesting problem in the space-air-ground (SAG) digital twin network. Thereof, digital twins operate as the tasks that need to be executed instantly at edge servers, indicating a lack of continuity across time slots and the potential flexibility achieved by twin migration.
Furthermore, in \cite{DT_blockchain_8}, the authors devised a dynamic algorithm for resource allocation of DNN partition empowered industrial internet of things (IoTs). Similar to \cite{DT_blockchain_7}, the absence of twin migration and task continuity makes the network static. In addition, the pre-set digital twin association in the network further affects the effectiveness of resource utilization of digital twins.

To tackle these issues,
researchers investigated the mechanism by which DITENs operate on a long-term time scale.
For instance, the authors in \cite{DT_traditional_1} aimed to enhance the quality of service (QoS) and reduce system costs of service migration tasks in the DITEN by optimizing service association and resource allocation. A traffic prediction-enabled multi-agent federated DRL approach was utilized to solve this problem. Besides, it was demonstrated in \cite{DT_vehicular4} that an evolutionary strategy-based method effectively solved the task offloading problem in the DITEN, achieving multiple objectives such as minimizing delay and energy costs while maximizing QoS. In addition, a task offloading problem was formulated in a vehicular edge computing network in \cite{DT_vehicular1}. Then, the authors proposed a generative adversarial network-based prediction method to reduce the number of parameters required to be optimized. Although these studies formulated the long-term system in the DITEN, the digital twins therein were still in the conceptual stages without detailed data structures. Consequently, evaluating the influence of digital twins on enhancing system efficiency was challenging in these papers.

Based on this consideration, other works attempted to formulate the digital twin structures in the DITEN to investigate the benefits of digital twins for the systems.
In \cite{DT_vehicular3}, the QoS of data sensing and data synchronization in the DITEN was optimized by adjusting digital twin associations and allocating sensing frequencies. Subsequently, a multi-agent DRL-based algorithm was adopted to solve this optimization problem.
Regarding latency minimization, an important aspect of QoS management, in \cite{DT_traditional_2}, the authors utilized the DITEN to minimize task offloading latency in an MEC system while satisfying the maximum latency requirements of each task. They employed a DRL-based algorithm to find solutions by optimizing offloading decisions.
Moreover, a network selection problem for data synchronization in the DITEN was proposed in \cite{DT_vehicular2}. The authors first used DRL to predict the waiting time for synchronization to minimize synchronization delay. Subsequently, they adopted a game-theoretic-based approach to solve the problem.
In \cite{DT_migration}, an edge association problem and a placement problem were formulated to minimize latency in the DITEN. A DRL-based algorithm was employed to solve the association problem, while a transfer learning-based algorithm was used to address the placement problem. The aforementioned works investigated digital twin placement, synchronization, and capacity estimation, and their impacts on DITENs. However, these digital twins are located at edge servers or cloudlets, making privacy considerations important. Unfortunately, these works overlooked this aspect.


Based on the discussions above, it can be noted that the joint consideration of both digital twin privacy and long-term operation in DITEN was overlooked by researchers. Many of them typically focused on one side. This may decrease the practicability of the DITEN system implementation.
To address this challenge, this paper focuses on jointly considering the optimization of digital twin-empowered FL tasks and the long-term operation of the DITEN.
Specifically, we formulate an optimization problem for the digital twin association and the historical data allocation used to enhance training performance in the DITEN. This performance is evaluated based on the data utility, which is measured with the assistance of digital twins.
Our objective is to balance the data utility of digital twin-empowered FL tasks and the energy costs associated with the long-term training procedure and maintenance of the DITEN. To tackle this multi-faceted optimization problem, we propose an optimization-driven learning algorithm that aims to find effective solutions. The main contributions of this paper can be summarized as follows:

\begin{itemize}

  \item By addressing concerns related to users' computing ability and the network's privacy leakage, we investigate the FL implementation on the DITEN. Capitalizing on the advantages of DITEN, we introduce a closed-form utility function to predict training accuracy for FL procedures, referred to as the data utility. Subsequently, we formulate a maximization problem to balance the benefits of FL on the DITEN and its associated energy costs. By optimizing the user digital twin association and historical data allocation, it maximizes the data utility while minimizing energy costs.

  \item The formulated optimization problem is a long-term, non-convex mixed-integer programming problem, posing significant challenges for finding the optimal solution. Specifically, balancing the benefits of data utility with the incurred energy cost is challenging. This challenge also arises from how to fully consider the trade-off between energy cost reduction from twin synchronization and the energy cost associated with twin migration. For tractability, an optimization-driven learning algorithm is explored.

  \item We explore an optimization-driven learning algorithm to address the above challenges. Specifically, we first decouple the historical data allocation problem from the original problem and prove that this decoupled problem is convex. Subsequently, we reformulate the association problem as a Markov decision process (MDP) and employ an online DRL-based algorithm to solve it. Numerical simulation results show the superiority compared to various baselines.
\end{itemize}

\begin{table*}
  \footnotesize
  \centering
  {\caption{List of Notations}}
  \begin{tabular}{@{}ll||ll@{}}
    \toprule
    Notation & Definition & Notation & Definition \\
    \midrule
    $S$                                     & Number of edge servers        & $U$                                     & Number of users\\
    $\cS$                                   & Index set of edge servers     & $\cU$                                   & Index set of users\\
    $T$                                     & Number of time slots          & $\cT$                                   & Index set of time slots\\
    $\cU_{s}^{t}$                           & Index set of users whose digital twins are established in $s$ at $t$ & $\text{DT}_{u}^{t}$ & Digital twin of $u$ at $t$\\
    $\omega^{t}_{u}$                        & Model weights of $u$ at $t$   & $\bm{\omega}^{t}$                       & Set of personalized model weights at $t$\\
    $\cD_{u}^{\text{twin}}$                 & Twin base data of $u$         & $\cD_{u,t}^{\text{dat}}$                & Training data of $u$ at $t$\\
    $D_{u}^{\text{twin}}$                   & Data size of twin base data for $u$ & $D_{u,t}^{\text{dat}}$            & Data size of training data for $u$\\
    $g_{s,u}^{t}$                           & Status information of $u\in\cU_{s}^{t}$ at $t$ & $p_{s}$                & Communication resource of $s$\\
    $q_{s}$                                 & Computation capacity of $s$   & $m_{s,u}^{t}$                           & Location of $u\in\cU_{s}^{t}$ at $t$\\
    $\bm{\kappa}^{t}$                       & Indicator set for the association results of user digital twins at $t$ & $\phi_{u}$ & EMD metric of $u$\\
    $\kappa_{u,s}^{t}$                      & Indicator for the association result between $u$ and $s$ at $t$ & $\bm{\gamma}^{t}$ & Set of historical data allocation ratios at $t$\\
    $\cD^{\text{dat}}_{u,t,t-1}$            & Historical training data for $u$ at $t$ & $\bar{\cD}^{\text{dat}}_{u,t}$ & Total training data for $u$ at $t$\\
    $\bar{D}^{\text{dat}}_{u,t}$            & Data size of total training data for $u$ at $t$ & $\cD^{\text{dat}}_{s,t}$ & Training data at $s$ in $t$\\
    $D^{\text{dat}}_{s,t}$                  & Data size of training data at $s$ in $t$ & $\cD_{t}^{\text{dat}}$       & Total training data at $t$\\
    $D_{t}^{\text{dat}}$                    & Data size of total training data at $t$ & $\omega^{t}$                  & Global training model weights at $t$\\
    $\omega_{s}^{t}$                        & Local training model weights for $s$ at $t$ & $\mathbb{P}_{u}$          & Distribution of training data for $u$\\
    $\mathbb{P}_{\text{all}}$               & Distribution of total training data & $\rho_{u,t}$                      & Data utility of $u$ at $t$\\
    $\alpha_{1}-\alpha_{6}$                 & Fitting coefficients          & $C_{s,t}^{\text{mig}}$                  & Migration costs of $s$ at $t$\\
    $n_{\text{mig}}$                        & Unit energy cost of wired transmission & $l_{s,s'}$ & Manhattan distance between $s$ and $s'$\\
    $n_{\text{syn}}$                        & Unit energy cost of wireless transmission & $C_{s,t}^{\text{syn}}$ & Synchronization cost of $s$ at $t$\\
    $l_{u,s}^{t}$                           & Manhattan distance between $u$ and $s$ at $t$ & $W$                     & Bandwidth of subchannels\\
    $P_{u}$                                 & Transmitting power of $u$     & $\xi$                                   & Channel power gain of subchannels\\
    $N_{0}$                                 & Noise power                   & $C_{s,t}^{\text{cmp}}$                  & Computation cost of $s$ at $t$\\
    $n_{\text{cmp}}$                        & Energy cost of one CPU cycle  & $\varepsilon_{s}$                       & CPU cycles required to compute one unit of data at $s$\\
    $E_{s}$                                 & Number of epoches for local training at $s$ & $E_{u,s}$                 & Number of epoches for fine-tuning for $u$ at $s$\\
    $\chi$                                  & Learning rate of the FL       & $C_{s,t}$                               & Total cost of $s$ at $t$\\
    $e^{t}$                                 & Objective value at $t$        & $\beta_{1},\beta_{2}$                   & Weight coefficient of the objective function\\
    $f_{0}$                                 & Factor of the normalized function & $a[t]$                              & Action at $t$\\
    $s[t]$                                  & State at $t$                  & $r[t]$                                  & Reward at $t$                 \\
    $\epsilon$                              & Factor of variation restricting for the actor network & $f$             & Curve coefficient of the barrier function\\
    $\pi_{\bm{\theta}}$                     & Association policy            & $\bm{\theta}$                           & Parameter vector of the actor network\\
    $\bm{\mu}$                              & Parameter vector of the critic network & $V_{\bm{\mu}}$                 & Value function of the critic network\\
    $\varphi_{t}$                           & Advantage between episodes at $t$ & $\sigma$                            & Discount factor of the advantage\\
    $\lambda$                               & Balance factor of the advantage & $A^{\text{GAE}}_{t}$                  & Advantage function\\
    $M$                                     & Number of epochs of the optimization-driven learning algorithm & $L^{\text{Act}}$ & Loss function of the actor network\\
    $\chi_{c}$                              & Learning rate of the critic network& $L^{\text{Cri}}$                   & Loss function of the critic network\\
    $\chi_{a}$                              & Learning rate of the actor network & & \\
    \bottomrule
  \end{tabular}
  \label{tab1}
\end{table*}

The rest of this paper is organized as follows. Section \ref{sec2} provides an overview of the DITEN system. In Section \ref{sec3}, we elaborate on the FL task in the DITEN, discussing its convergence property and presenting the problem formulation for the digital twin association and the historical data allocation. Section \ref{sec4} introduces our proposed optimization-driven learning algorithm to solve the problem. Section \ref{sec5} presents the experimental setup and numerical results. Finally, Section \ref{sec6} concludes this paper and indicates the future research directions.
The notations used throughout the paper are summarized in Table I.

\section{System model}
\label{sec2}

This section presents our proposed DITEN system, which utilizes digital twins to assist in personalized model training.
We begin by introducing the components of the DITEN. Then, we elaborate on the details of the FL training procedure. Afterward, an evaluation metric of training performance is given. Finally, we delve into the associated energy costs involved in this procedure.

\subsection{System Model}
We consider a DITEN comprising one cloud server, $S$ edge servers, and $U$ users.
Let $\cS = \{1,2,...,S\}$ denote the index set of all edge servers, and $\cU = \{1,2,...,U\}$ denote the index set of all users. In this network, two types of digital twins are established: user digital twins and edge server digital twins, as illustrated in Fig. \ref{fig1}. Due to the limited computational and storage capabilities, users do not have the ability to establish their digital twins independently \cite{DT_independ1,DT_independ2}.
Therefore, during time slot $t\in\cT=\{1,2,...,T\}$, user digital twins are established in edge servers. For a specific edge server $s\in\cS$, we define $\cU_{s}^{t}\subset\cU$ as the set of users whose digital twins are currently established in this edge server. In addition, for a user $u\in\cU_{s}^{t}$, its digital twin can be expressed as
\begin{equation}
\text{DT}_{u}^{t} = \{\omega^{t}_{u},\cD_{u}^{\text{twin}},\cD_{u,t}^{\text{dat}}\},
\end{equation}
where $\omega^{t}_{u}$ represents the current model weights of $u$, $\cD_{u}^{\text{twin}}$ is the base twin data of user $u$, which represents the fundamental programs and data required for user $u$'s digital twin implementation. $\cD_{u,t}^{\text{dat}}$ depicts the current training data of user $u$ that is collected and updated by the user through uploads in each time slot. The sizes of two types of data are denoted by $D_{u}^{\text{twin}} = |\cD_{u}^{\text{twin}}|$ and $D_{u,t}^{\text{dat}} = |\cD_{u,t}^{\text{dat}}|$, respectively, where $|\cdot|$ is the cardinality operator for set. Correspondingly, as for edge server $s$, its digital twin in time slot $t$ is described as
\begin{equation}
\text{DT}_{s}^{t} = \{g_{s,u}^{t},p_{s},q_{s}\},
\end{equation}
where $g_{s,u}^{t} = \{m_{s,u}^{t},\phi_{u},D_{u}^{\text{twin}},D_{u,t}^{\text{dat}}\}$
represents the status information of user $u\in\cU_{s}^{t}$.
In addition, the status information includes the user's current location $m_{s,u}^{t}\in\mathbb{R}^{2}$, the earth mover's distance (EMD) metric $\phi_{u}$ which measures dissimilarity between two frequency distributions and will be detailed in Section \ref{subsec2}, the base twin data size $D_{u}^{\text{twin}}$, and the current training data size $D_{u,t}^{\text{dat}}$. Moreover, $p_{s}$ and $q_{s}$ represent the communication resource and the computational capacity of edge server $s$, respectively. Furthermore, this type of digital twin is established in the cloud server. Both digital twins are models that map the physical status of their corresponding entities and are updated in each time slot. Based on the established edge server digital twins, the cloud server can monitor the status of the DITEN and guide the association between user digital twins and edge servers, as mentioned above. To describe the digital twin associations, we introduce $\bm{\kappa}^{t} = \{\kappa_{u,s}^{t}:u\in\cU,s\in\cS\}$ as the indicator set for the association states of user digital twins at time slot $t$. Specifically, $\kappa_{u,s}^{t} = 1$ means that user $u$ establishes its digital twin in edge server $s$ at time slot $t$, and $\kappa_{u,s}^{t} = 0$ otherwise.
Based on the association result,
details regarding the model training is provided in the following subsection.

\subsection{Federated Learning Model}
\label{subsec1}

\begin{figure}
    \centering
    \includegraphics[width=2.5in]{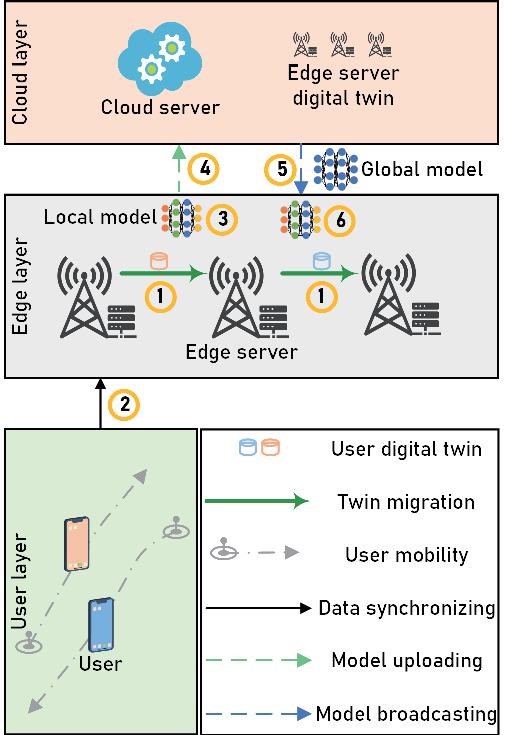}
    \caption{System model of the DITEN.}
    \label{fig1}
\end{figure}

In DITEN, privacy is a significant concern, as personalized data is continuously synchronized between digital twins and their physical entities \cite{DT_blockchain_1}. To address this concern, in this subsection, FL is leveraged to execute model training tasks in the DITEN, preventing data leakage to the cloud server or other users.
In detail, users and the cloud server cannot access the entire dataset. The reason to prevent data processing in the cloud server is twofold: the heavy traffic burden caused by the distance between the cloud server and users and the increased vulnerability to malicious attacks compared to edge servers \cite{DT_traditional_3}. 
Specifically,
by utilizing advanced privacy-preserving authentication technologies \cite{privacy_pre3}, edge servers and the cloud server can be considered trustworthy by users. Therefore, privacy risks in servers typically originate from third parties, such as eavesdroppers \cite{privacy_pre4}.
Compared to the cloud server, edge servers exhibit a higher frequency of changes in stored data, making personalized data more challenging to trace and attack.
Therefore, edge servers can only access user data associated with themselves, and neither users nor the cloud server can access the complete dataset in our DITEN. During the data transmission, developed privacy preserving technologies \cite{privacy_pre1,privacy_pre2} can be adopted to further ensure the privacy.
Besides this, to enhance the quality of training results, we allow for the inclusion of full or partial historical data from the previous time slot $t-1$ in the training process at the current time slot $t$. Denote $\bm{\gamma}^{t} = \{\gamma_{u}^{t}:u\in\cU\}$ as the set for inclusion ratios at $t$. In specific, $\gamma_{u}^{t}\in[0,1]$ represents the ratio of user $u$'s training data from time slot $t-1$ that is included in the training at time slot $t$. Correspondingly, we define $\cD^{\text{dat}}_{u,t,t-1}$ as a subset of $\cD^{\text{dat}}_{u,t-1}$ with a cardinality equal to $\gamma_{u}^{t}\cdot D^{\text{dat}}_{u,t-1}$. The distribution of this subset matches that of $\cD^{\text{dat}}_{u,t-1}$.
Therefore, the dataset that $\text{DT}_{u}^{t}$ is used for FL at time slot $t$ is the union of $\cD^{\text{dat}}_{u,t}$ and $\cD^{\text{dat}}_{u,t,t-1}$. For the sake of conciseness, we define this dataset as
\begin{equation}
\bar{\cD}^{\text{dat}}_{u,t} = \cD^{\text{dat}}_{u,t}\cup\cD^{\text{dat}}_{u,t,t-1},
\end{equation}
and define its size as $\bar{D}^{\text{dat}}_{u,t}$. Moreover, regarding the aspect of an edge server, the training dataset that edge server $s$ uses at time slot $t$ is the union of the datasets its user digital twins use. It can be defined as $\cD^{\text{dat}}_{s,t} = \{\bar{\cD}^{\text{dat}}_{u,t}:u\in\cU,\kappa_{u,s}^{t} = 1\}$. We define its size as $D^{\text{dat}}_{s,t} = |\cD^{\text{dat}}_{s,t}|$. In addition, all the data that participates in the aggregation at time slot $t$ can be expressed as $\cD_{t}^{\text{dat}}=\cup_{s\in\cS}\cD_{s,t}^{\text{dat}}$ with a size of $D_{t}^{\text{dat}} = |\cD_{t}^{\text{dat}}|$. Based on these preliminaries, we provide more detailed FL procedures and their loss functions.

As mentioned previously, FL aims to establish personalized model weights $\omega^{t}_{u}$, $\forall u$. This process involves six steps, as illustrated in Fig. \ref{fig1}. In step $1$, user digital twins migrate to target edge servers based on the current associated results $\bm{\kappa}^{t}$. Subsequently, users synchronize with their respective digital twins in step $2$. Following this, edge servers train their respective model weights $\omega^{t}_{s}$, using $\cD^{\text{dat}}_{s,t}$, where $s\in\cS$, as shown in step $3$. In step 4, to enhance the robustness of the trained model, the global model weights $\omega^{t}$ are aggregated at the cloud server as follows:
\begin{equation}
\omega^{t} = \frac{1}{D_{t}^{\text{dat}}}\sum_{s\in\cS}D_{s,t}^{\text{dat}}\cdot \omega_{s}^{t}.
\end{equation}
Subsequently, $\omega^{t}$ is broadcasted to all edge servers in step $5$. Steps $1$-$5$ are repeated multiple times until the model converges. In step $6$, the personalized model weights $\omega^{t}_{u}$ for $u\in\cU$ are generated through fine-tuning at edge servers using user-specific data ($\bar{\cD}^{\text{dat}}_{u,t}$, $\forall u$).
The entire training procedure is assumed to be completed within the time slot $t$. Based on these training procedures, we can provide loss functions for our FL tasks. Specifically, let's define $\{x,y\}$ as a sample of training data and denote the loss function as $f(\omega,x,y)$, which quantifies the difference between estimated and true values for instances of running data under any model weights $\omega$. Correspondingly, the loss function of the local training in step $3$
is defined as
\begin{equation}
F(\omega_{s}^{t}) = \frac{1}{D_{s,t}^{\text{dat}}}\sum_{\{x,y\}\in\cD_{s,t}^{\text{dat}}} f(\omega_{s}^{t},x,y).
\end{equation}
Subsequently, the aggregated loss function of the cloud server in the step $4$ is given by
\begin{equation}
F(\omega^{t}) = \frac{1}{D_{t}^{\text{dat}}}\sum_{\{x,y\}\in\cD_{t}^{\text{dat}}} f(\omega^{t},x,y).
\end{equation}
The global loss is minimized as follows:
\begin{equation}
\underset{\omega^{t}}{\min}~F(\omega^{t}).
\label{weightOpt}
\end{equation}
Afterward, in the step $6$, the loss function for fine-tuning is represented as
\begin{equation}
F(\omega_{u}^{t}) = \frac{1}{\bar{D}^{\text{dat}}_{u,t}}\sum_{\{x,y\}\in\bar{\cD}^{\text{dat}}_{u,t}} f(\omega_{u}^{t},x,y).
\end{equation}
In this step, the goal is to minimize the following fine-tuning loss:
\begin{equation}
\underset{\bm{\omega}^{t}}{\min}\sum_{u\in\cU}F(\omega^{t}_{u}),
\end{equation}
where $\bm{\omega}^{t} = \{\omega^{t}_{u}:u\in\cU\}$.

It should be noted that the cloud server should accomplish the digital twin association and the historical data allocation at the beginning of each training round to achieve the FL procedure. However, critical challenges arise when the cloud server directly associates digital twins and allocates data. To be more specific,
based on the discussion at the beginning of this subsection,
it is evident that the cloud server does not have to access the original training data. Instead, it only possesses status information $g_{s,u}^{t}$, $\forall u,s,t$. As a result, the cloud server faces challenges in predicting the training accuracy when each user digital twin is used for FL and thus struggles to estimate the potential contributions that each user digital twin provides. Consequently, the association of appropriate edge servers for user digital twins and the historical data allocation of user digital twins become difficult. The difficulty stems from the inability to balance the contributions (accurate model weights) that each user's digital twin can offer and the energy costs it incurs. To overcome this limitation, we provide a performance evaluation metric in the next subsection to allow the cloud server to proceed with the association and allocation process before this FL procedure. Furthermore, in Section \ref{sec5}, the actual training result will be compared with the performance evaluation to assess the effectiveness of this evaluating method.

\subsection{Data Utility Model}
\label{subsec2}

This subsection proposes an evaluation method that assesses the data quality of user digital twins in FL training by introducing the concept of data utility for user digital twins. Moreover, to allow the cloud server to obtain this data quality evaluation, the method only utilizes status information ($g_{s,u}^{t}$) to compose the data utility of each user's digital twin.
By using $g_{s,u}^{t}$, it is straightforward to obtain the EMD metric $\phi_{u}$ and training data size $\bar{D}^{\text{dat}}_{u,t}$, which are used to establish the proposed data utility of $\text{DT}_{u}^{t}$. Besides, due to the method such as parametric dual bound functions can estimate $\phi_{u}$ for each user with error guarantee \cite{EMD_classic,EMD_guarantee}, we assume that the EMD metric for each user digital twin is trusted. Therefore, we introduce the EMD preliminaries at first and then formulate the data utility. In FL scenarios, data distribution is typically non-identically distributed (non-i.i.d). The divergence of data distribution between local devices and the overall dataset is one of the main reasons for decreased FL accuracy \cite{nonIID}. This divergence can be measured by the EMD metric. Specifically, we consider an $L$-classification task whose overall dataset is $\cX$ and its label set is $\cY$, which follows the distribution $\mathbb{P}_{\text{all}}$.
Due to the non-i.i.d nature of the data distribution in the FL procedure, the distribution of the dataset in $\text{DT}_{u}^{t}$, denoted by $\mathbb{P}_{u}$, differs from that of $\cX$. That is, the proportion of the data class labeled by $y\in\cY$ in $\mathbb{P}_{u}$ is different from that in $\mathbb{P}_{\text{all}}$. Based on this context, EMD between user $u$ and the overall dataset can be defined by
\begin{equation}
\phi_{u} = \sum_{y\in\cY}||\mathbb{P}_{u}(y) - \mathbb{P}_{\text{all}}(y)||,
\end{equation}
where $\mathbb{P}_{u}(y)$ and $\mathbb{P}_{\text{all}}(y)$ represent the proportion of the data sample labeled $y$ in $\bar{\cD}^{\text{dat}}_{u,t}$ and the proportion of overall dataset, respectively. $||\cdot||$ represents L1 norm.

According to the empirical formula provided by \cite{utilityCurve}, the data utility, which can be depicted as the training accuracy prediction of $u$ in this task, can be expressed as $\rho_{u,t}\in[0,1]$, calculated by
\begin{equation}
\rho_{u,t} = \upsilon(\phi_{u}) - \alpha_{1}\cdot e^{-\alpha_{2}\cdot(\alpha_{3}\cdot \bar{D}^{\text{dat}}_{u,t})^{\upsilon(\phi_{u})}},
\label{utility}
\end{equation}
in which,
\begin{equation}
\upsilon(\phi_{u}) = \alpha_{4} \cdot e^{-(\frac{\alpha_{5}+\phi_{u}}{\alpha_{6}})^{2}},
\end{equation}
where $\alpha_{k}>0$ $(1\leq k \leq 6)$ are fitting coefficients. In this formula, $\upsilon(\phi_{u})$ indicates that a larger EMD is more detrimental to the learning performance. On the other hand, the exponential part $\alpha_{1}\cdot e^{-\alpha_{2}\cdot(\alpha_{3}\cdot \bar{D}^{\text{dat}}_{u,t})^{\upsilon(\phi_{u})}}$ indicates that a larger data size will reduce this minus term, thereby benefiting the learning accuracy.

\subsection{Energy Consumption Model}
\label{subsec3}

Based on the previous discussions, we comprehensively analyze the inherent energy costs associated with the training process in this subsection.
These energy costs include digital twin migration, training data synchronization, and model computation costs.

\subsubsection{Migration energy cost}

When a user's digital twin association changes from time slot $t-1$ to $t$, this digital twin needs to be migrated to the target edge server from the original edge server, incurring corresponding energy costs. We assume that each pair of edge servers is connected by fibers without relays. Therefore, the migration energy cost regarding target edge server $s$ at time slot $t$ is only related to the data size that needs to be migrated and the distance between it and the original edge server. This energy cost can be depicted as
\begin{equation}
C_{s,t}^{\text{mig}} = n_{\text{mig}} \cdot \sum_{s'\in\cS}\sum_{u\in\cU} l_{s,s'} \cdot (\bar{D}^{\text{dat}}_{u,t} + D_{u}^{\text{twin}}) \cdot \kappa_{u,s}^{t} \cdot \kappa_{u,s'}^{t - 1},
\end{equation}
where $n_{\text{mig}}$ is the energy cost of transmitting one unit of data over one unit of distance in wired transmission. In addition, $l_{s,s'}$ is the Manhattan distance \cite{Manhattan_1,Manhattan_2} between edge server $s$ and $s'$ where $s,s'\in\cS$ in the two-dimensional scenario.

\subsubsection{Synchronization energy cost}

In our DITEN, users generate personalized data at time slot $t$ and synchronize by uploading their data to the corresponding associated edge servers after all migrations are accomplished. We consider a multi-cast orthogonal frequency-division multiple access (OFDMA) protocol between users and edge servers in the DITEN. Assuming that all users use their transmitting power $P_{u}$ to upload data, and all subchannels have a bandwidth of $W$, according to \cite{DT_user_selection_2}, the synchronization energy cost for edge server $s$ at time slot $t$ is expressed as follows:
\begin{equation}
C_{s,t}^{\text{syn}} = n_{\text{syn}}\cdot\sum_{u\in\cU}\frac{D_{u,t}^{\text{dat}}}{W\cdot\log_{2}(1+\frac{P_{u}\cdot\xi}{N_{0}})}\cdot l_{u,s}^{t}\cdot \kappa_{u,s}^{t},
\end{equation}
where $n_{\text{syn}}$ represents the energy cost of transmitting one unit of synchronizing time over one unit of distance in wireless transmission, $l_{u,s}^{t}$ is the Manhattan distance between user $u$ and edge server $s$ at time slot $t$. In addition, $\xi$ is each subchannel's unified channel power gain, and $N_{0}$ is the noise power.

\subsubsection{Computation energy cost}

After the digital twin migration and the training data synchronization, the DITEN trains personalized models for all users. As mentioned, our FL training consists of local training at edge servers, global aggregation at the cloud server, and fine-tuning at edge servers. It is worth noting that the training data size is much larger than the size of models in FL. Therefore, the energy costs of transmitting the model between edge servers and the cloud server, and those among edge servers are negligible. Moreover, the cloud server has sufficient computational resources to handle the model aggregation, which only requires linear time complexity about the edge server number. Therefore, we only consider the energy costs of training at edge servers. This part of energy costs consists of the energy cost incurred by local training and the energy cost generated by fine-tuning for each user digital twin.
We assume edge server $s$ needs $E_{s}$ epochs to train its local model. Correspondingly, for user $u\in\cU_{s}^{t}$, its digital twin is assumed to use $E_{u,s}$ epochs to accomplish the fine-tuning procedure.
Let $\varepsilon_{s}$ be the CPU cycles required to compute one unit of data at edge server $s$. For each edge server, this type of energy cost can be described as \cite{DT_user_selection_2}
\begin{equation}
C_{s,t}^{\text{cmp}} = n_{\text{cmp}}\cdot\varepsilon_{s}\cdot\sum_{u\in\cU}\bar{D}^{\text{dat}}_{u,t}\cdot(E_{s} + E_{u,s})\cdot \kappa_{u,s}^{t},
\end{equation}
where $n_{\text{cmp}}$ is the energy cost of one CPU cycle.
Based on the system description and incurred energy costs, in the next section, we will first demonstrate the convergence of the training process we proposed.
Subsequently, we formulate an optimization problem to balance data utility and energy costs.

\section{FL Convergence Analysis and Problem Formulation}
\label{sec3}

In this section, we will initially analyze the convergence of the FL training procedure for the DITEN.
Then, we formulate an optimization problem to maximize data utility while minimizing energy costs.

\subsection{FL Convergence Analysis}

Based on our DITEN system, various training scenarios can be applied. These scenarios typically involve different FL algorithms, and different loss functions are defined accordingly \cite{ConvergeThou}. Examples of such loss functions include mean squared error, logistic regression, and cross-entropy.
Without loss of generality, we define a specific FL algorithm. In this algorithm, each training round is formed by one time of local gradient descent for each edge server, and one time of aggregation for the cloud server.
In the step $3$ and $4$ of the FL procedure, as mentioned in Section \ref{subsec1}, the global weights start from $\omega^{t,0}$, iterate through several rounds, and converge to the optimal solution of (\ref{weightOpt}), denoted as $\omega^{t,*}$. Correspondingly, we define the weights in the $i$-th round as $\omega^{t,i}$. Similarly, in the step $6$, we define the weights of user $u$ in $i$-th round as $\omega^{t,i}_{u}$ and define its optimal solution as $\omega^{t,*}_{u}$.
To ensure the convergence of the FL procedure in these scenarios, we first make some assumptions about the FL procedure and then provide a convergence analysis under these assumptions, similar to the approach outlined in \cite{ConvergeTemplate}.
\begin{itemize}
  \item First, we assume that the gradient of loss function for the cloud server, which is denoted by $\nabla F(\omega^{t,i})$, is uniformly Lipschitz continuous concerning $\omega^{t,i}$ \cite{ConvergeLip}. Therefore, we have
      \begin{equation}
      ||\nabla F(\omega^{t,i+1})-\nabla F(\omega^{t,i})|| \leq L\cdot||\omega^{t,i+1}-\omega^{t,i}||,
      \end{equation}
      where $L>0$ is correlated to the loss functions.
  \item Second, we assume that $F(\omega^{t,i})$ is strongly convex with $\psi>0$, such that
      \begin{align}
      F(\omega^{t,i+1})\leq F(\omega^{t,i}) + & (\omega^{t,i+1}-\omega^{t,i})^{\text{T}}\nabla F(\omega^{t,i})\nonumber\\ &+\frac{\psi}{2}||\omega^{t,i+1}-\omega^{t,i}||^{2}.
      \end{align}
  \item Third, we assume that $F(\omega^{t,i})$ is twice-continuously differentiable. Therefore, we have
      \begin{align}
      \psi\bm{I}\preceq\nabla^{2}F(\omega^{t,i})\preceq L\bm{I},
      \end{align}
      where $\bm{I}$ is the identity matrix.
\end{itemize}
Based on these assumptions, the convergence rate of the FL algorithm can be obtained by the following theorem.
\begin{myth}
\label{ConvergeTheorem}
Given the learning rate $\chi$, global optimal weights $\omega^{t,*}$, and fine-tuned optimal weights $\omega^{t,*}_{u}$, $\forall u$, the upper bound of $F(\omega^{t,i+1}) - F(\omega^{t,*})$ and $F(\omega^{t,i+1}_{u}) - F(\omega^{t,*}_{u})$ can be given by
\begin{align}
&F(\omega^{t,i+1}) - F(\omega^{t,*})\leq \nonumber\\
&e^{-(i+1)(2\psi\chi-L\psi\chi^{2})}(F(\omega^{t,0}) - F(\omega^{t,*})),
\end{align}
and
\begin{align}
&F(\omega^{t,i+1}_{u}) - F(\omega^{t,*}_{u})\leq \nonumber\\
&e^{-(i+1)(2\psi\chi-L\psi\chi^{2})}(F(\omega^{t,0}_{u}) - F(\omega^{t,*}_{u})), \forall u,
\end{align}
respectively.
\end{myth}
\begin{IEEEproof}
Please refer to Appendix \ref{ProofConverge}.
\end{IEEEproof}

\subsection{Problem Formulation}

The objective function of the optimization problem consists of the data utility and energy costs. The energy costs are determined by factors such as the digital twin migration, the training data synchronization, and the model computation. Therefore, the energy costs for edge server $s$ at time slot $t$ are depicted as follows:
\begin{equation}
C_{s,t} = C_{s,t}^{\text{mig}} + C_{s,t}^{\text{syn}} + C_{s,t}^{\text{cmp}}.
\label{total_cost}
\end{equation}
Our considered objective function at time slot $t$ is given as follows:
\begin{equation}
e^{t}(\bm{\kappa}^{t},\bm{\gamma}^{t}) = \frac{\beta_{1}}{U}\cdot\sum_{u\in\cU}\rho_{u,t} - \frac{\beta_{2}}{S}\cdot\sum_{s\in\cS} \text{Norm}(C_{s,t}, f_{0}),
\end{equation}
where $\beta_{1}$, $\beta_{2}\in(0,1]$ are weight coefficients that balance data utility and energy costs, and $\text{Norm}(x, f_{0}) = \frac{2}{1+e^{-\frac{x}{2f_{0}}}}-1$ is a normalized function \cite{Norm} that can normalize the energy costs into $[0,1]$ based on a scaling factor $f_{0}$.
Consequently, the optimization problem can be formulated as a long-term mixed-integer programming problem, which is expressed as follows:
\begin{align}
\label{P1}\underset{\bm{\kappa}^{t},\bm{\gamma}^{t}}{\max}~ & \frac{1}{T}\sum_{t\in\cT}e^{t}(\bm{\kappa}^{t},\bm{\gamma}^{t})\\
\mathrm{ s.t.}~ & \mathrm{C1:}~ \sum_{s\in\cS} \kappa_{u,s}^{t} = 1, ~\forall u,t,\nonumber\\
& \mathrm{C2:}~ C_{s,t}^{\text{cmp}} \leq q_{s}, ~\forall s,t,\nonumber\\
& \mathrm{C3:}~ C_{s,t}^{\text{syn}} \leq p_{s}, ~\forall s,t,\nonumber\\
& \mathrm{C4:}~ \kappa_{u,s}^{t}\in\{0,1\}, ~\forall u,s,t,\nonumber\\
& \mathrm{C5:}~ \gamma_{u}^{t}\in[0,1], ~\forall u,t\nonumber,
\end{align}
where C1 illustrates that each user digital twin can only select one edge server at time slot $t$. C2 ensures that the computation energy consumption of each edge server does not surpass its computational capacity. C3 represents that the communication resources used in the synchronization procedure in each edge server can not exceed its limit. C4 and C5 depict the integrity requirements for the optimization variables.
It is noteworthy that (\ref{P1}) is a long-term mixed-integer programming problem. By optimizing user digital twin association, the system resource can be fully utilized. Correspondingly, more historical data can be employed to enhance the data utility, thereby elevating the FL training performance.
Due to the non-convex and nonlinear nature of the proposed problem, as well as the dynamic variants and the vast solution space, solving this optimization problem using traditional methods such as the alternating direction method of multipliers \cite{ADAM} or the block successive upper-bound minimization \cite{BSUM} becomes extremely challenging. To address these issues, we present an optimization-driven learning algorithm designed to effectively and efficiently find high-quality solutions for the given problem in the following section.

\section{Algorithm Design}
\label{sec4}

In this section, we propose an optimization-driven learning algorithm to solve (\ref{P1}).
Specifically, we first reformulate the original problem into an MDP and then provide
with details of the algorithm to solve this MDP.
In general, this algorithm solves the user digital twin association result $\bm{\kappa}^{t}$ using a DRL-based method while addressing the historical data allocation $\bm{\gamma}^{t}$ through standard convex optimization. The optimization method we used relies on the learning result of $\bm{\kappa}^{t}$ to obtain the solution of $\bm{\gamma}^{t}$, which further
guides the learning process of the association results in the next time slot. The solution to (\ref{P1}) is obtained through this iterative approach.

\subsection{MDP Reformulation}
\begin{figure}
    \centering
    \includegraphics[width=2.2in]{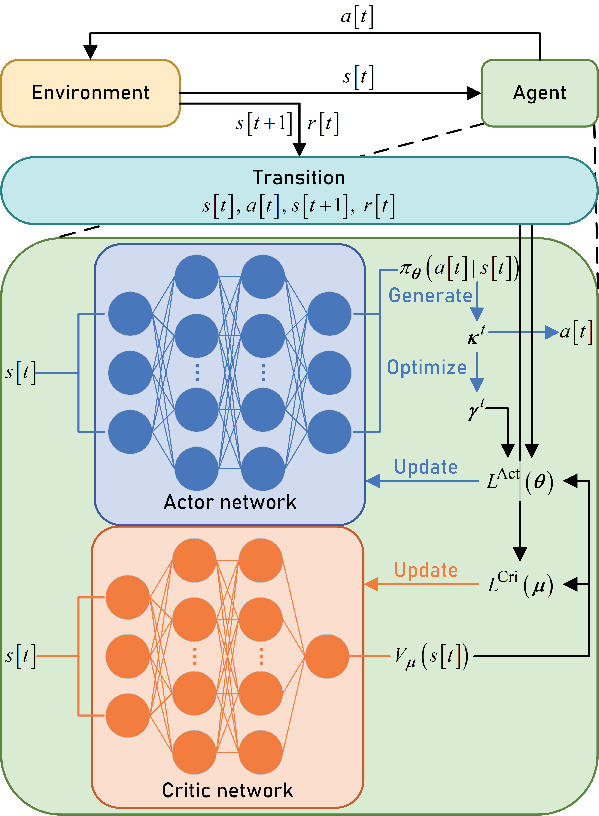}
    \caption{An illustration of our optimization-driven learning algorithm.}
    \label{fig2}
\end{figure}

In this subsection, we reformulate (\ref{P1}) as an MDP to solve it using the optimization-driven learning algorithm. Aligning with the time slots of (\ref{P1}), in our problem, the user digital twin association and the historical data allocation are performed in $T$ episodes, where each episode's operation is only influenced by the previous episode. During episode $t\in\cT$, an agent deployed on the cloud server generates an action $a[t]$ based on the current state $s[t]$ and receives a reward $r[t]$ from the environment. Consequently, the agent repeatedly collects transitions in the form of $\{s[t],~a[t],~s[t+1],~r[t]\}$ for learning purposes, as depicted at the top of Fig. \ref{fig2}. Building upon this framework, we proceed to provide detailed definitions of state $s[t]$, action $a[t]$, and reward $r[t]$.
\subsubsection{State}
To provide feedback on the current scenario information to the cloud server, the state of the MDP includes the status information of user digital twins, the locations of edge servers, and the distances between users and the associated edge server digital twins. It can be defined as $s[t] \triangleq \{(g_{s,u}^{t},m_{s},l_{u,s}^{t}):s\in\cS,u\in\cU_{s}^{t}\}$, where $m_{s}$ is the location of edge server $s$.

\subsubsection{Action}
Since we will solve the historical data allocation $\bm{\gamma}^{t}$ using the learning result of the user digital twin association, the action of the MDP in the current episode only includes the latter. It can be depicted as $a[t] \triangleq \bm{\kappa}^{t}$.

\subsubsection{Reward}
After the agent proceeds with actions at episode $t$, it will obtain the corresponding reward from the environment. This reward needs to reflect both the objective and constraints of the optimization problem. Therefore, we define a logarithmic barrier function to combine the objective function and constraints. This barrier function is depicted as
\begin{equation}
\label{barrier}
\cB(x) =
\begin{cases}
-\frac{1}{f}\ln(-x), & x<0, \\
\infty , & x \geq 0,
\end{cases}
\end{equation}
where $f>0$ is a curve coefficient. By using this barrier function,
the reward of the MDP can be defined as
\begin{equation}
r[t] \triangleq e^{t}(\bm{\kappa}^{t},\bm{\gamma}^{t}) - \sum_{s\in\cS}\cB(C_{s,t}^{\text{cmp}}-q_{s}) - \sum_{s\in\cS}\cB(C_{s,t}^{\text{syn}}-p_{s}).
\label{reward}
\end{equation}
As mentioned earlier, the algorithm combines a DRL-based and a standard optimization method. To illustrate how the optimization method interacts with the learning process, we first introduce the convex optimization method, assuming that $\bm{\kappa}^{t}$ is given in episode $t$. We then use the optimized $\bm{\gamma}^{t}$ to learn $\bm{\kappa}^{t}$.

\subsection{Optimization of Data Allocation}
\label{sec4subsec1}

At episode $t$,
the user digital twin association result from the previous episode $\bm{\kappa}^{t-1}$ is known,
we also assume that $\bm{\kappa}^{t}$ is already given.
Consequently, the optimization of the historical data allocation in episode $t$ can be decoupled from (\ref{P1}) and formulated as follows:
\begin{align}
\label{P2}\underset{\bm{\gamma}^{t}}{\min}~ & -e^{t}(\bm{\gamma}^{t})\\
\mathrm{ s.t.}~ & \mathrm{C1^{\dag}:}~ C_{s,t}^{\text{cmp}} - q_{s}\leq 0, ~\forall s,\nonumber\\
& \mathrm{C2^{\dag}:}~ -\gamma_{u}^{t}\leq 0, ~\forall u\nonumber,\\
& \mathrm{C3^{\dag}:}~ \gamma_{u}^{t} - 1\leq 0, ~\forall u\nonumber.
\end{align}
\begin{myth}
\label{TheConcave}
Problem (\ref{P2}) is convex.
\end{myth}
\begin{IEEEproof}
In (\ref{P2}), all $\gamma_{u}^{t}$ ($u\in\cU$) are continuous real variables. Thus, $-e^{t}(\bm{\gamma}^{t})$ is second-order differentiable, and its second-order derivative is quoted below
\begin{equation}
-\frac{\partial^{2}e^{t}(\bm{\gamma}^{t})}{\partial\gamma_{u}^{t}\partial\gamma_{v}^{t}} =
\begin{cases}
-\eta_{1}(\gamma_{u}^{t})\cdot\eta_{2}(\gamma_{u}^{t})+\eta_{3}(\gamma_{u}^{t}), & \text{if}~u = v \\
0, & \text{otherwise},
\label{secDiff}
\end{cases}
\end{equation}
where
\begin{equation}
\eta_{1}(\gamma_{u}^{t}) = \frac{(\alpha_{3}D_{u,t}^{\text{dat}})^{2}\alpha_{1}\alpha_{2}\beta_{1}\upsilon(\phi_{u})\cdot e^{-\alpha_{2}(\alpha_{3}\bar{D}^{\text{dat}}_{u,t})^{\upsilon(\phi_{u})}}}{U\cdot(\alpha_{3}\bar{D}^{\text{dat}}_{u,t})^{2-\upsilon(\phi_{u})}},
\label{secDiffApp1}
\end{equation}
\begin{equation}
\eta_{2}(\gamma_{u}^{t}) = \upsilon(\phi_{u})-1-\alpha_{2}\upsilon(\phi_{u})(\alpha_{3}\bar{D}^{\text{dat}}_{u,t})^{\upsilon(\phi_{u})},
\label{secDiffApp2}
\end{equation}
and
\begin{equation}
\eta_{3}(\gamma_{u}^{t}) = \sum_{s\in\cS}\frac{\beta_{2}e^{-\frac{C_{s,t}}{f_{0}}\cdot(1-e^{-\frac{C_{s,t}}{2f_{0}}})}}{S\cdot f_{0}^{2}(e^{-\frac{C_{s,t}}{2f_{0}}}+1)^{3}}\cdot \bar{C}_{s,t}^{2},
\label{secDiffApp3}
\end{equation}
where $\bar{C}_{s,t}$ is given by
\begin{equation}
\bar{C}_{s,t} = D_{u,t-1}^{\text{dat}}\kappa_{u,s}^{t}(\sum_{s'\in\cS}n_{\text{mig}}l_{s,s'}\kappa_{u,s'}^{t-1}+n_{\text{cmp}}\varepsilon_{s}(E_{s}+E_{u,s})).
\label{secDiffApp4}
\end{equation}
Thereof, $\eta_{1}(\gamma_{u}^{t})$ and $\eta_{3}(\gamma_{u}^{t})$ are large than 0. For $\eta_{2}(\gamma_{u}^{t})$, $0<\upsilon(\phi_{u})<1$ and $\alpha_{2}\upsilon(\phi_{u})(\alpha_{3}\bar{D}^{\text{dat}}_{u,t})^{\upsilon(\phi_{u})}>0$. Thus, from (\ref{secDiff}), we can get that $-\frac{\partial^{2}e^{t}(\bm{\gamma}^{t})}{\partial\gamma_{u}^{t}\partial\gamma_{v}^{t}}\geq 0$ ($u\in\cU$), so the Hessian matrix of $-e^{t}(\bm{\gamma}^{t})$ is diagonal and the elements on the diagonal are all non-negative, and $-e^{t}(\bm{\gamma}^{t})$ is a convex function. In addition, the constraints of problem (\ref{P2}) are all linear.
This completes the proof \cite{alterAlgoTWC,alterAlgoOrig}.
\end{IEEEproof}
Then, utilizing conventional convex optimization \cite{NewtonQi}, the optimal solution of (\ref{P2}) can be obtained directly.
In the next subsection, with the obtained $\bm{\gamma}^{t}$, we will use the DRL-based learning process to solve $\bm{\kappa}^{t}$.

\subsection{DRL-based Learning Algorithm}

As mentioned earlier, the learning process of the algorithm is carried out by an agent located at the cloud server. Therefore, we first introduce the structure of the agent and then demonstrate the detailed learning process.
The design of the agent is based on that of the proximal policy optimization (PPO) algorithm \cite{PPO}. Compared to other DRL algorithms such as actor-critic algorithm \cite{actorCritic} and deep deterministic policy gradient (DDPG) \cite{DDPG}, PPO ensures better monotonicity of learning performance and can effectively learn a randomness policy.
Specifically, the agent consists of two networks: the actor network and the critic network, as depicted at the bottom of Fig. \ref{fig2}. The actor network is responsible for learning the association strategy $\pi_{\bm{\theta}}(a[t]|s[t])$, while the critic network learns the value of the executed action to adjust the strategy learning. Similar to the parameter vector of the actor network $\bm{\theta}$, we define the parameter vector of the critic network as $\bm{\mu}$. Accordingly, we define the value function of the state as $V_{\bm{\mu}}:\mathbb{R}^{d}\rightarrow\mathbb{R}$, where $d$ is the dimension of the state.

Based on the introduced agent, the learning process can be achieved by learning the objective function for the actor network and the loss function for the critic network in parallel. Specifically, for the actor network, we utilize generalized advantage estimation (GAE) to find the advantages that each episode gets. We then define the objective function using the PPO-clip method \cite{PPO}. By defining a discount factor $\sigma$, we can represent the advantage between episodes as follows:
\begin{equation}
\varphi_{t} = r[t]+\sigma\cdot V_{\bm{\mu}}(s[t+1])-V_{\bm{\mu}}(s[t]).
\end{equation}
Afterward, defining a parameter $\lambda\in[0,1]$ to balance short-term and long-term advantages, the advantage function during the learning process can be defined as
\begin{equation}
A^{\text{GAE}}_{t} = (1-\lambda)\cdot\sum_{j = 1}^{\infty}\sum_{l=0}^{j-1}\lambda^{j-1}\cdot\sigma^{l}\cdot\varphi_{t+l}.
\label{A_GAE}
\end{equation}
We assume that the training process iterates $M$ epochs to converge. Based on this advantage function, for $m$-th epoch of the iteration, the parameter $\bm{\theta}_{m}$ can be updated by maximizing the objective function (\ref{LAct}).
\begin{figure*}[!t]
\normalsize
\begin{equation}
L^{\text{Act}}(\bm{\theta}) = \mathbb{E}\left[\min(\frac{\pi_{\bm{\theta}_{m}}(a[t]|s[t])}{\pi_{\bm{\theta}_{1}}(a[t]|s[t])}\cdot A^{\text{GAE}}_{t},
\text{Clip}(\frac{\pi_{\bm{\theta}_{m}}(a[t]|s[t])}{\pi_{\bm{\theta}_{1}}(a[t]|s[t])},1-\epsilon,1+\epsilon)\cdot A^{\text{GAE}}_{t})\right].
\label{LAct}
\end{equation}
\end{figure*}
Thereof, $\mathbb{E}[\cdot]$ is expectation, $\epsilon$ is the factor that restrict the variation of ${\bm{\theta}}_{m}$ in each epoch by using the clip function defined by
\begin{equation}
\text{Clip}(a,b,c) \triangleq \max(\min(a,b),c).
\end{equation}
On the other hand, for the critic network, we define a minimized mean squared error (MSE) based loss function for its updating. This function can be depicted as
\begin{equation}
L^{\text{Cri}}(\bm{\mu}) = \frac{1}{2}\varphi_{t}^{2}.
\end{equation}
Based on the objective function and loss function introduced above, the actor network and the critic network can be updated by the gradient ascent and the gradient descent, respectively. Defining the learning rates $\chi_{a}$ and $\chi_{c}$, the updating formula are shown as follows:
\begin{equation}
\bm{\theta}_{m+1} = \bm{\theta}_{m} + \chi_{a}\cdot\nabla L^{\text{Act}}(\bm{\theta}_{m}),
\label{UptAct}
\end{equation}
\begin{equation}
\bm{\mu}_{m+1} = \bm{\mu}_{m} - \chi_{c}\cdot\nabla L^{\text{Cri}}(\bm{\mu}_{m}).
\label{UptCri}
\end{equation}
Based on the discussions above, assuming the
strategy learning
repeats $P$ times, we can summarise the optimization-driven learning algorithm in the Algorithm \ref{alg3}.
\begin{algorithm}
	\caption{PPO-based digital twin association algorithm.}
	\label{alg3}
	\begin{algorithmic}[1]
            \STATE Determine $\sigma$, $\lambda$, $\epsilon$, $\chi_{a}$, $\chi_{c}$.
            \STATE Initialize network parameters $\bm{\theta}$, $\bm{\mu}$.
            \FOR{$p=1,...,P$}
                \FOR{$t=1,...,T$}\label{linePart1Beg}
                    \STATE Use the actor network to calculate $\pi_{\bm{\theta}}(a[t]|s[t])$.
                    \STATE Use the critic network to calculate $V_{\bm{\mu}}(s[t])$.
                    \STATE Generate $\bm{\kappa}^{t}$ based on the policy $\pi_{\bm{\theta}}(a[t]|s[t])$.
                    \STATE Optimize $\bm{\gamma}^{t}$ using convex optimization.
                    \STATE Calculate (\ref{reward}) to get $r[t]$.
                    \STATE Update $s[t+1]$.
                    \STATE Collect transition $\{s[t],~a[t],~r[t],~s[t+1]\}$.
                \ENDFOR\label{linePart1End}
                \FOR{$m=1,...,M$}\label{linePart2Beg}
                    \STATE Calculate $L^{\text{Act}}(\bm{\theta})$ and $L^{\text{Cri}}(\bm{\mu})$.
                    \STATE Update $\bm{\theta}$ and $\bm{\mu}$ using (\ref{UptAct}) and (\ref{UptCri}), respectively.
                \ENDFOR\label{linePart2End}
            \ENDFOR
            \RETURN Strategy parameters $\bm{\theta}$.
	\end{algorithmic}
\end{algorithm}
The learning algorithm consists of a main loop that can be divided into two parts. The first part of the algorithm (line \ref{linePart1Beg}-\ref{linePart1End}) describes the interaction between the agent and the environment. The agent first constructs the user digital twin association result based on the current policy and then optimizes the historical data allocation result to propose action. Subsequently, the reward is calculated based on the action, and the transitions of each episode are collected for the second part.
The second part of the algorithm (line \ref{linePart2Beg}-\ref{linePart2End}) describes the policy update based on the collected transitions ($\{s[t],~a[t],~r[t],~s[t+1]\}$, $\forall t$) from the first part. Transitions obtained from association and allocation results are used as historical data for the network's learning.
We then determine the computational complexity of the optimization-driven learning algorithm. We utilize the multi-layer perceptron (MLP) in neural networks. Denoting $H_{k}$ as the number of neurons in the $k$-th layer, the computational complexity of a $K$-layer MLP for one step is given by $\cO(\sum_{k=2}^{K-1}H_{k-1}H_{k}+H_{k}H_{k+1})$. The actor and critic networks are based on MLP. Furthermore, we employ the interior point method (IPM) \cite{NewtonQi} for convex optimization in the algorithm. Define $I_{t,p}$ and $J_{t,p}$ as the iteration times for step size searching and gradient descent at episode $t$ in $p$-th strategy learning, respectively. Hence, the computational complexity of the convex optimization is given by $\cO(\sum_{t=1}^{T}\sum_{p=1}^{P}I_{t,p}J_{t,p})$. Therefore, the overall computational complexity is calculated as $\cO(P(M+T)\sum_{k=2}^{K-1}(H_{k-1}H_{k}+H_{k}H_{k+1})+\sum_{t=1}^{T}\sum_{p=1}^{P}I_{t,p}J_{t,p})$.
In the next section, numerical results will demonstrate the performance and efficiency of the proposed optimization-driven learning algorithm.

\section{Numerical Results}
\label{sec5}

In this section, we conduct extensive numerical simulations to validate the performance of our proposed algorithm. It is important to note that the development of DITENs is still in its early stages, and there are currently no publicly available workload traces specifically designed for DITENs. Therefore, in our simulations, we generate synthetic instances to address the problems of the user digital twin association and the historical data allocation. We consider a DITEN deployed in a two-dimensional area of size $120\times120~m^{2}$ area, where a certain number of edge servers and users are uniformly distributed throughout this region \cite{settingArea}.
Specifically, we fix the number of users at $20$, while varying the number of edge servers from $9$ to $21$.
For each user digital twin, in the absence of workload trace, we refer to the training data size in \cite{DT_user_selection_1} to determine the size of the twin base data. This size ranges from $4.4$ Kb to $4.6$ Kb.
The training data size is given by the actual size of the MNIST\footnote{http://yann.lecun.com/exdb/mnist/} dataset.

\begin{figure}
    \centering
    \includegraphics[width=3in]{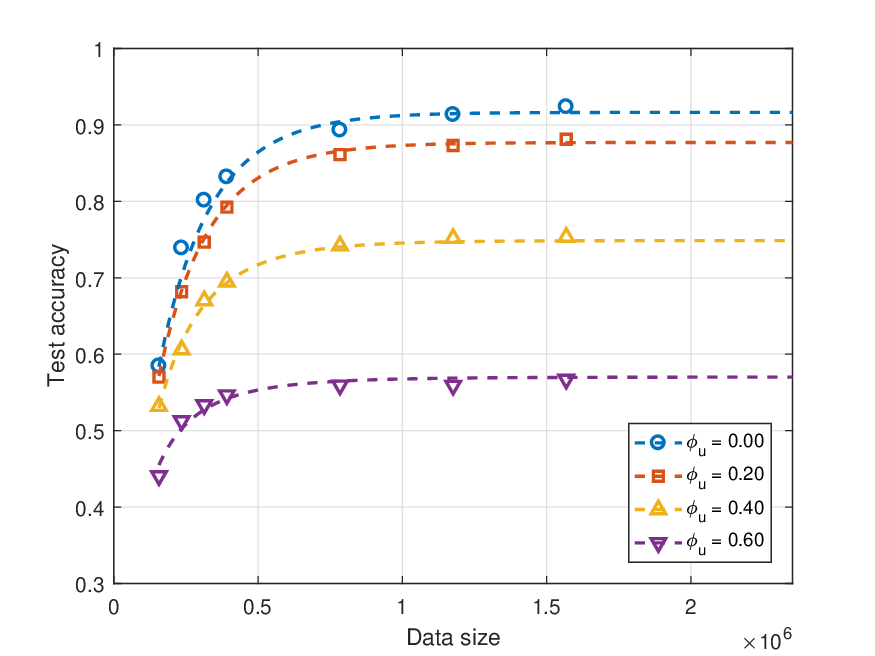}
    \caption{Estimation of the data utility function in (\ref{utility}).}
    \label{fig3}
\end{figure}

Moreover,
by jointly considering representativeness and redundancy avoidance,
we set four different EMD values: $\phi_{u} = 0.00$, $\phi_{u} = 0.20$, $\phi_{u} = 0.40$, and $\phi_{u} = 0.60$, $\forall u$.
Specifically, the range of the settings is referenced in \cite{EMDuse}, and the intervals are detailed in \cite{utilityCurve,DT_user_selection_2}.
We employ ResNet-18 \cite{ResNet} with a learning rate of $\chi=0.007$ for FL training. Furthermore, we perform association and allocation over $750$ time slots in each specific scenario instance, i.e., $T=750$. Within each time slot, every edge server undergoes $50$ rounds of training and $10$ rounds of fine-tuning for users \cite{utilityCurve}. Based on these experimental settings, we vary the data size $\bar{D}^{\text{dat}}_{u,t}$ and EMD metric $\phi_{u}$ for user $u$ to demonstrate the learning accuracy trends in our proposed FL procedure. We estimate this trend using the data utility introduced in (\ref{utility}) to showcase the effectiveness of our model. As depicted in Fig. \ref{fig3}, the number of user image samples ranges from $200$ to $2000$, while each MNIST dataset image remains fixed at $28\times 28$ bits. Consequently, the $\bar{D}^{\text{dat}}_{u,t}$ varies from $1.57\times10^{5}$ to $1.57\times10^{6}$. The scattered spots in the figure represent training accuracy evaluated using the aforementioned experimental settings. At the same time, the dashed lines indicate data utilities estimated by (\ref{utility}) with parameters $\alpha_{1}$ to $\alpha_{6}$ derived from the training accuracy. For our experiments, we set the following values for the parameters $\alpha_{1}$ to $\alpha_{6}$: $\alpha_{1}=0.8862$, $\alpha_{2}=6.8382$, $\alpha_{3}=0.0006$, $\alpha_{4}=0.9172$, $\alpha_{5}=-0.0231$ and $\alpha_{6}=0.8366$ in the following experiments.

Regarding the energy consumption model of the scenarios, the experimental parameters are mainly referenced from \cite{DT_traditional_3,DT_user_selection_1,utilityCurve,PPO}.
Specifically, communication resources range from $120$ to $150$, while computational capacities of edge servers vary from $1.4\times10^{3}$ to $1.5\times10^{3}$, following a uniform distribution. In other words, $p_{s}\in U[120,150]$ and $q_{s}\in U[1.4,1.5]\times10^{3}$, $\forall s$.
In addition, we set the required CPU cycles for one unit of data $\varepsilon_{s}$ to range from $54$ to $56$. The unit wired transmission energy cost, the unit wireless transmission energy cost, and the unit computation energy cost are set as $10^{-4}$, $10^{-1}$, and $10^{-7}$, respectively. For details of the synchronization energy cost, we set the users' transmission power as $200$ mW, and their bandwidth as $15$ KHz. The unified channel power gain is $1$, and the noise power is $-174$ dBm. Finally, the normalized factor $f_{s}$ is set as $200$, and $\beta_{1}$ and $\beta_{2}$ are set to $0.3$ and $0.7$ respectively.

As for the parameters of the proposed algorithm, we adopt a standard fully connected network for both the actor network and the critic network, with $128$ neurons for each hidden layer. The learning rate of the actor network  $\chi_{a}$ is $2.5\times10^{-4}$, while the learning rate of the critic network $\chi_{c}$ is $1.5\times10^{-3}$. The curve coefficient of the barrier function $f$ is set to $10$, and the discount factor $\sigma$, the balance factor $\lambda$, and the restrict factor $\epsilon$ are set as $0.98$, $0.9$, and $0.2$, respectively. The key parameters used throughout the simulation are summarized in Table \ref{Tab1}.
\begin{table}
    \footnotesize
    \centering
    \caption{Parameter settings}
    \scalebox{0.9}{ \begin{tabular}{|c|c|c|}
        \hline
        {\bf Parameters}&\multicolumn{1}{c|}{\bf numerical values}\\ \hline
        The number of users $U$                                         & $20$                    \\ \hline
        The number of edge server $S$                                            & $9-21$                        \\ \hline
        The size of twin base data $D_{u}^{\text{twin}}$                & $U[4.4,4.6]\times10^{3}$      \\ \hline
        EMD value $\phi_{u}$                                          & $0.00$, $0.20$, $0.40$, $0.60$\\ \hline
        The FL learning rate $\chi$                                     & $0.007$                        \\ \hline
        The number of time slots $T$                                    & $750$                        \\ \hline
        Model training epochs $E_{s}$                                   & $50$                         \\ \hline
        Model fine tuning epochs $E_{u,s}$                              & $10$                          \\ \hline
        Fitting coefficients $\alpha_{1}$-$\alpha_{3}$                  & $0.8862$, $6.8382$, $0.0006$  \\ \hline
        Fitting coefficients $\alpha_{4}$-$\alpha_{6}$                  & $0.9172$, $-0.0231$, $0.8366$ \\ \hline
        Computation resources $q_{s}$                                   & $U[1.4,1.5]\times10^{3}$      \\ \hline
        Communication capacities $p_{s}$                                & $U[120,130]$                    \\ \hline
        Required CPU cycles $\varepsilon_{s}$                           & $U[54, 56]$                   \\ \hline
        The wired transmitting unit energy cost $n_{\text{mig}}$               & $10^{-4}$                     \\ \hline
        The wireless transmitting unit energy cost $n_{\text{syn}}$            & $10^{-1}$                     \\ \hline
        The computational unit energy cost $n_{\text{cmp}}$                    & $10^{-7}$                     \\ \hline
        User transmitting powers $P_{u}$                                & $200$ mW                      \\ \hline
        Bandwidth of users $W$                                          & $15$ KHz                      \\ \hline
        Channel power chain $\xi$                                       & $1$                           \\ \hline
        Noise power $N_{0}$                                             & $-174$ dBm                    \\ \hline
        Normalized factor $f_{s}$                                       & $200$                         \\ \hline
        Weight coefficient $\beta_{1}$                                  & $0.3$                         \\ \hline
        Weight coefficient $\beta_{2}$                                  & $0.7$                         \\ \hline
        Learning rate $\chi_{a}$, $\chi_{c}$                            & $2.5\times10^{-4}$, $1.5\times10^{-3}$ \\ \hline
        Curve coefficient $f$                                           & $10$                          \\ \hline
        Discount factor $\sigma$                                        & $0.98$                        \\ \hline
        Balance factor $\lambda$                                        & $0.9$                         \\ \hline
        Restrict factor $\epsilon$                                      & $0.2$                         \\ \hline
    \end{tabular}}
\label{Tab1}
\end{table}
For performance comparison, the following baselines are considered:
\begin{itemize}
  \item \textbf{Baseline 1}: This strategy employs an actor-critic network-based scheme \cite{actorCritic} to find solutions for user digital twin association while still employing the historical data allocation method from our proposed algorithm.

  \item \textbf{Baseline 2}: In this method, user digital twins are associated with the edge servers nearest to their physician entities while satisfying the given constraints. The historical data allocation scheme is identical to that of Baseline 1.

  \item \textbf{Baseline 3}: In this scheme, the user digital twin association method remains the same as that of Baseline 2. However, a random historical data allocation scheme is used.

  \item \textbf{Baseline 4}: This scheme adopts a deep deterministic policy gradient-based method \cite{DDPG} to solve the association problem, incorporating our proposed historical data allocation scheme.

\end{itemize}

\subsection{Convergence Performance}
\label{sec5a}

\begin{figure}
\centering
\subfigure[$\phi_{u} = 0.0$]{
\begin{minipage}[t]{0.45\linewidth}
    \centering
    \includegraphics[width=1\textwidth]{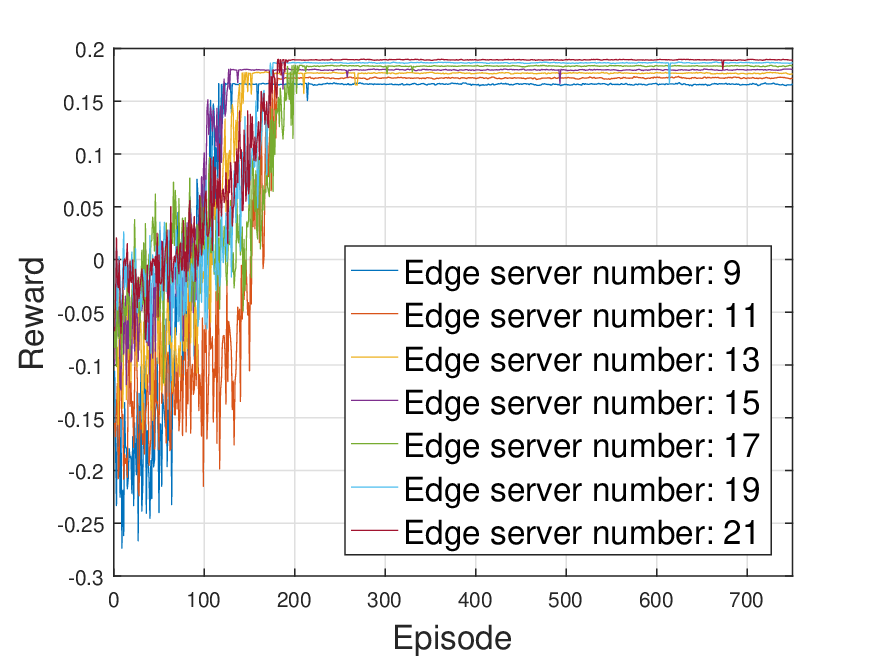}
\end{minipage}
}
\subfigure[$\phi_{u} = 0.2$]{
\begin{minipage}[t]{0.45\linewidth}
    \centering
    \includegraphics[width=1\textwidth]{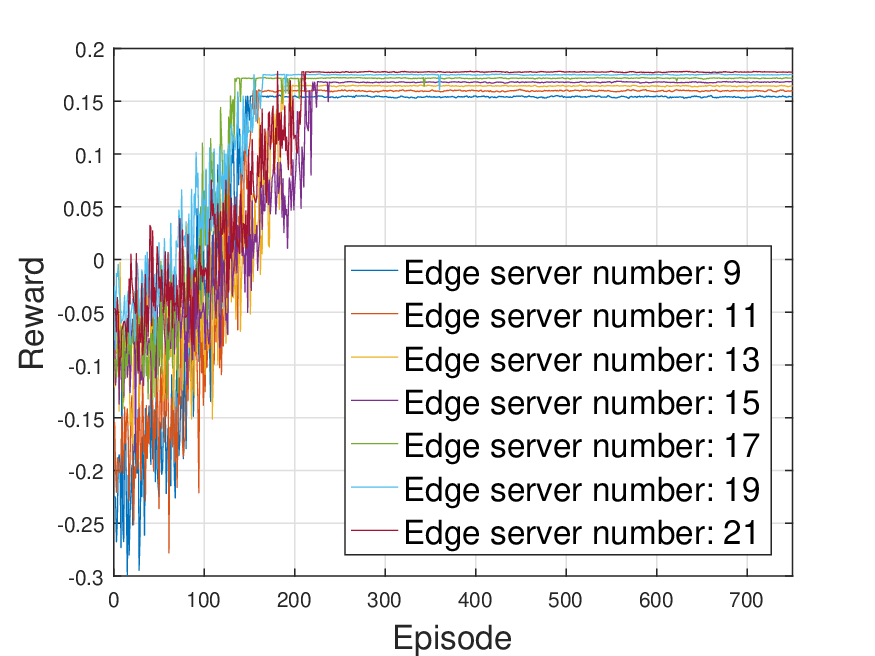}
\end{minipage}
}
\subfigure[$\phi_{u} = 0.4$]{
\begin{minipage}[t]{0.45\linewidth}
    \centering
    \includegraphics[width=1\textwidth]{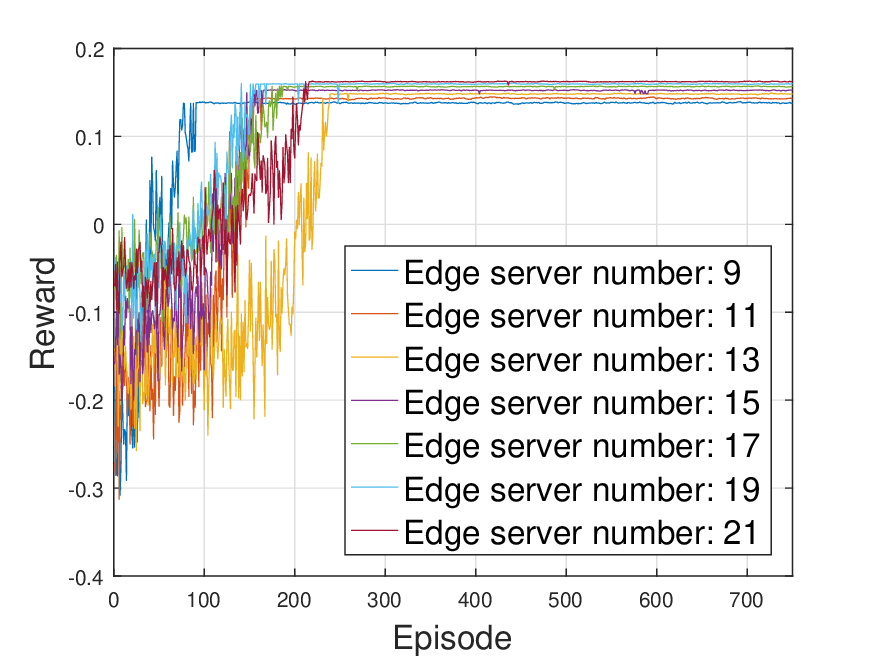}
\end{minipage}
}
\subfigure[$\phi_{u} = 0.6$]{
\begin{minipage}[t]{0.45\linewidth}
    \centering
    \includegraphics[width=1\textwidth]{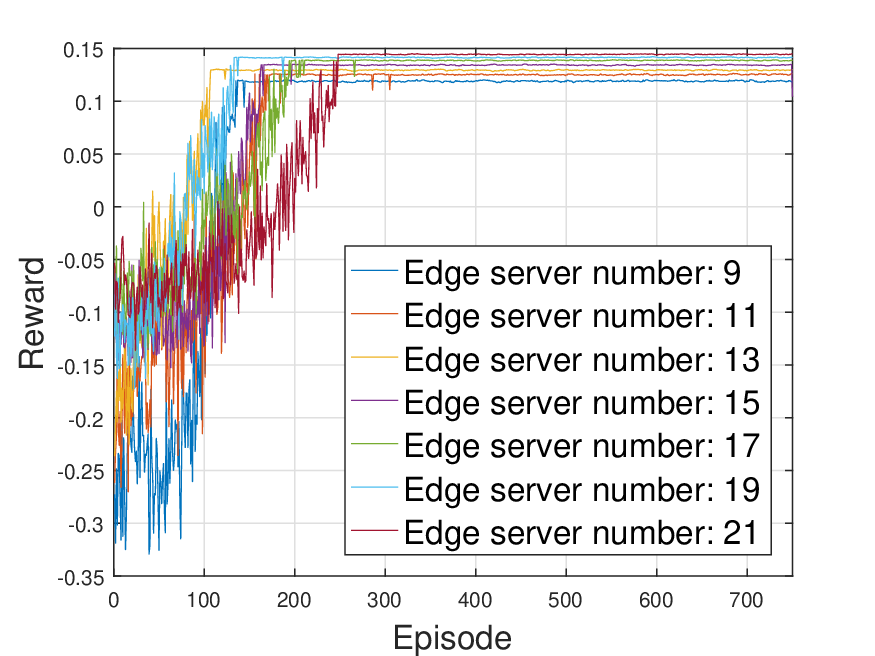}
\end{minipage}
}
\centering
\caption{Convergence performance of PPO algorithm with 20 users.}
\label{fig4}
\end{figure}

In this subsection, the convergence performance of our optimization-driven learning algorithm is evaluated under various parameter settings, as illustrated in Fig. \ref{fig4}.
Therein, the number of users is set to be $U=20$. In addition, the values of $\phi_{u}$ represent the EMD value indicating dataset distribution differences.
They are set to be $0.0$, $0.2$, $0.4$, and $0.6$ respectively in the Fig. \ref{fig4}(a), Fig. \ref{fig4}(b), Fig. \ref{fig4}(c), and Fig. \ref{fig4}(d).
We use the rewards during the episodes to show this performance. Specifically, the $x$-axis depicts the number of iterations while the $y$-axis represents the rewards in each episode calculated by (\ref{reward}).
Furthermore, to fully showcase the effects of different scenario settings on performance and avoid redundancy, we adopt scenario settings with varying numbers of edge servers: $9$, $11$, $13$, $15$, $17$, $19$, and $21$.

From each of the sub-figures, we can see that
our proposed algorithm
converges quickly under these settings.
In addition, after the rewards of the proposed algorithm reach their peak level, the rewards of each setting still fluctuate slightly due to user mobility and environmental variations.
Besides, as expected, the larger number of edge servers corresponds to higher
available resources, leading to higher rewards once convergence is achieved.
When comparing Fig. \ref{fig4}(a) to the other sub-figures, it can be observed that all the converged rewards of the respective edge server numbers in Fig. \ref{fig4}(a) are the highest among the sub-figures. These rewards then slightly decrease along with the growth of $\phi_{u}$ in Fig. \ref{fig4}(b), Fig. \ref{fig4}(c), and Fig. \ref{fig4}(d).
This implies that data with lower $\phi_{u}$ can achieve
better performance.

\subsection{Objective Value}
\label{sec5b}

\begin{figure}
\centering
\subfigure[$\phi_{u} = 0.0$]{
\begin{minipage}[t]{0.45\linewidth}
    \centering
    \includegraphics[width=1\textwidth]{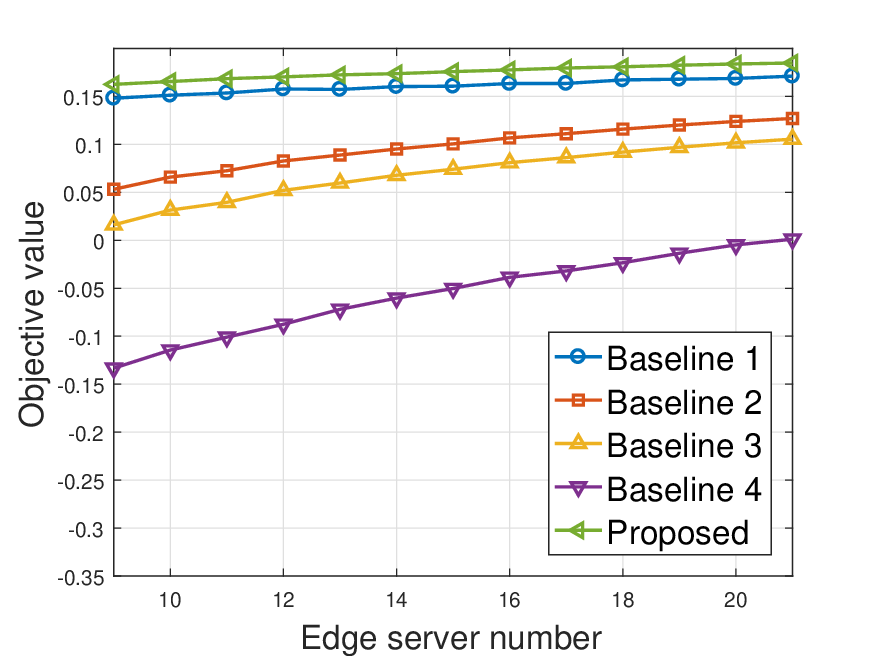}
\end{minipage}
}
\subfigure[$\phi_{u} = 0.2$]{
\begin{minipage}[t]{0.45\linewidth}
    \centering
    \includegraphics[width=1\textwidth]{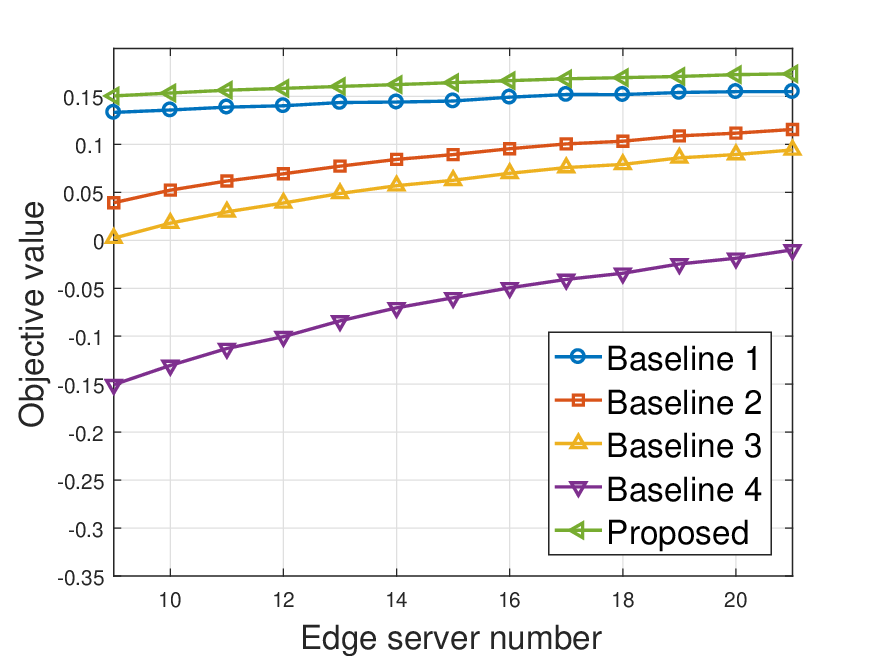}
\end{minipage}
}
\subfigure[$\phi_{u} = 0.4$]{
\begin{minipage}[t]{0.45\linewidth}
    \centering
    \includegraphics[width=1\textwidth]{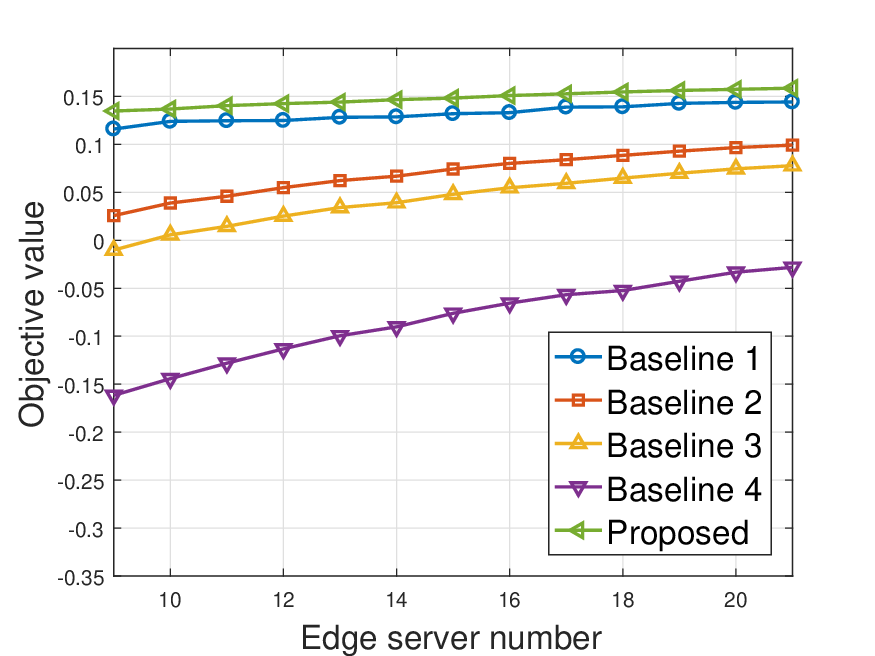}
\end{minipage}
}
\subfigure[$\phi_{u} = 0.6$]{
\begin{minipage}[t]{0.45\linewidth}
    \centering
    \includegraphics[width=1\textwidth]{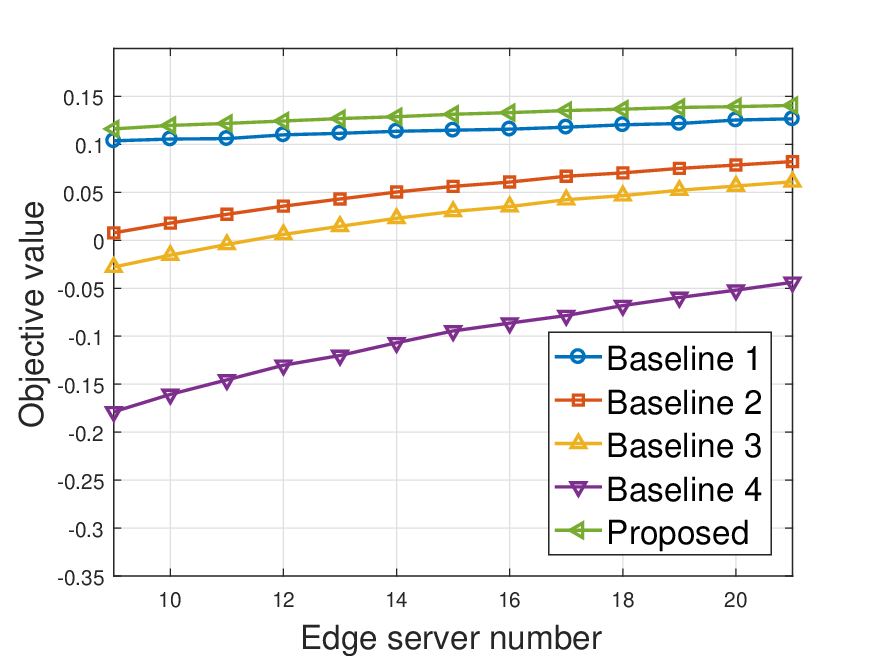}
\end{minipage}
}
\centering
\caption{Objective value versus edge server number for different algorithms with $20$ users.}
\label{fig5}
\end{figure}

Fig. \ref{fig5} illustrates the objective values of our proposed algorithm
and the extensive baselines versus the number of edge servers.
Similar to Section \ref{sec5a}, $\phi_{u}$ are set as $0.0$, $0.2$, $0.4$, and $0.6$, respectively, for the four sub-figures.
It is observed that the objective value for all of algorithms increase along with the increasing of the edge server numbers, indicating better performance brought by more available resources. In addition, the scale of performance increasing becomes slower when the number of edge servers is large, indicating that the resource is sufficient.
As expected, the objective value of all the schemes increases with the increase of the edge server number. In addition, our proposed algorithm outperforms the baselines for any given $\phi_{u}$.
Specifically, for these four sub-figures, our proposed algorithm achieves objective values that are, on average, $0.1022$, $0.5070$, $0.6842$, and $1.5025$ times higher than Baselines 1, 2, 3, and 4, respectively.
The reasons why our proposed algorithm surpasses the baselines can be summarized as follows. Compared to our proposed algorithm, Baseline 1's loss function does not effectively confine the search for solutions within a reasonable range in each iteration, leading to inferior convergence performance. As for Baselines 2 and 3, frequent user digital twin migrations would increase energy costs and impact system efficiency. Conversely, Baseline 4, due to the requirement of larger transitions, needs more episodes to converge, thereby decreasing its average performance. In contrast, our proposed algorithm utilizes a clipped loss function to control the search speed during iterations, enhancing performance compared to Baselines 1, 2, 3, and 4.

\subsection{Data Utility and Energy Costs}
\label{sec5c}

\begin{figure}[t]
    \centering
    \includegraphics[width=2.5in]{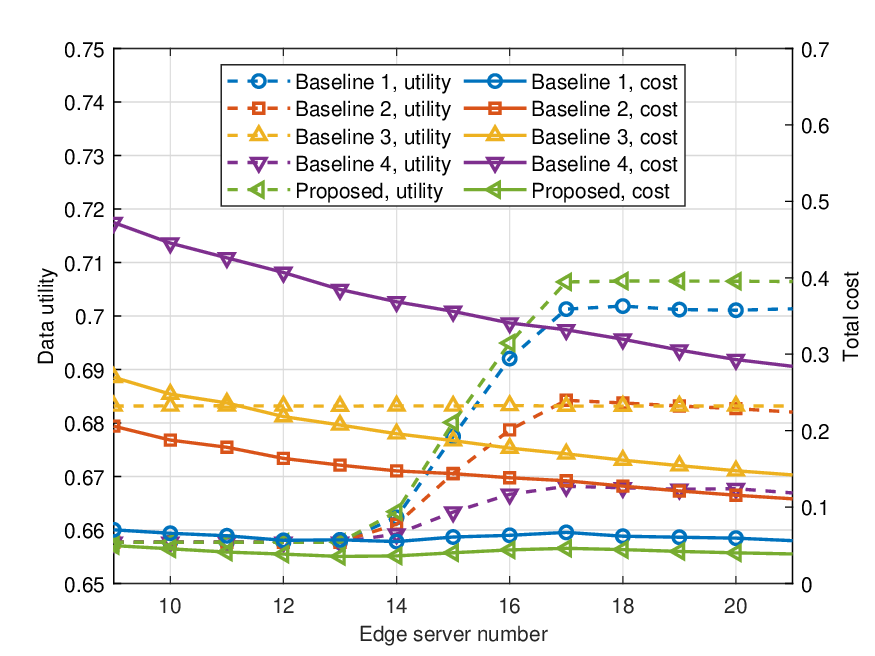}
    \caption{Data utilities and energy costs versus edge server number for different algorithms with $\phi_{u} = 0.0$.}
    \label{fig6}
\end{figure}

\begin{figure}[ht]
    \centering
    \includegraphics[width=2.5in]{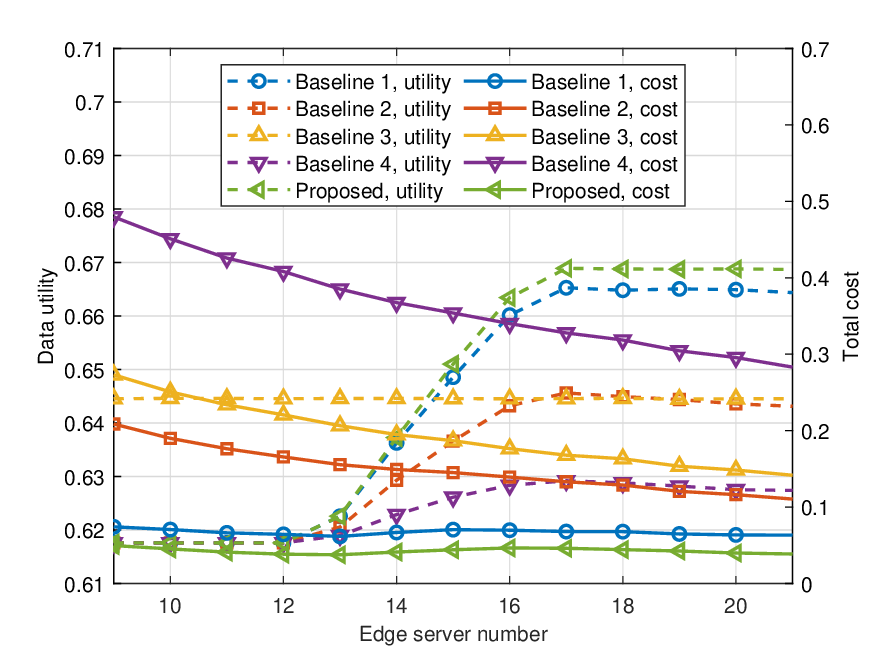}
    \caption{Data utilities and energy costs versus edge server number for different algorithms with $\phi_{u} = 0.2$.}
    \label{fig7}
\end{figure}

\begin{figure}[t]
    \centering
    \includegraphics[width=2.5in]{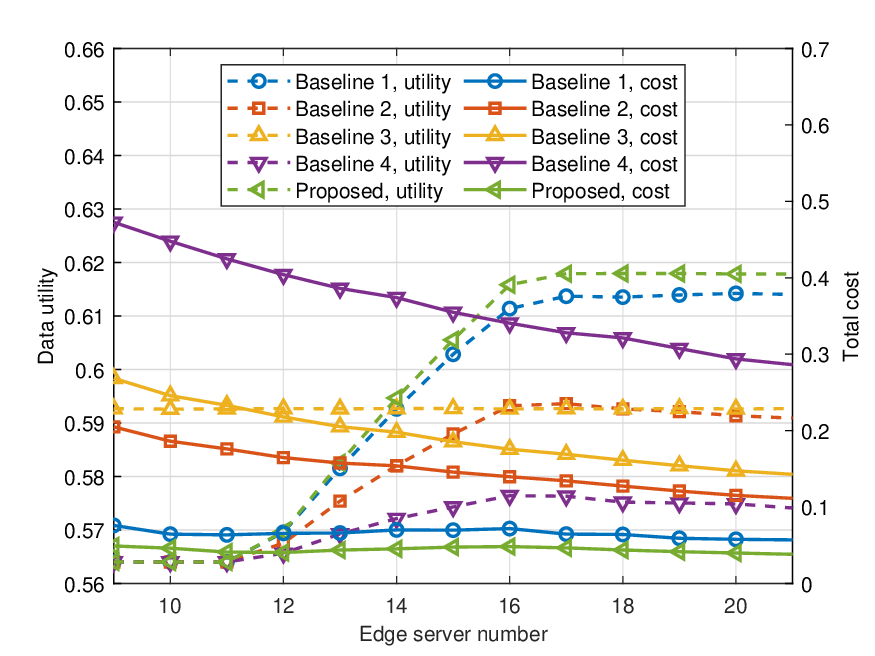}
    \caption{Data utilities and energy costs versus edge server number for different algorithms with $\phi_{u} = 0.4$.}
    \label{fig8}
\end{figure}

\begin{figure}[ht]
    \centering
    \includegraphics[width=2.5in]{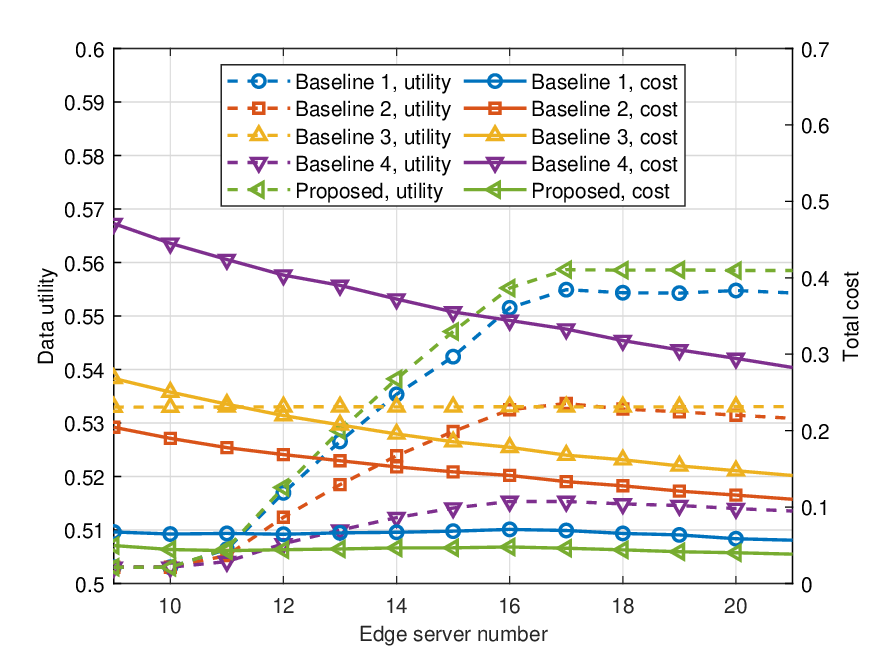}
    \caption{Data utilities and energy costs versus edge server number for different algorithms with $\phi_{u} = 0.6$.}
    \label{fig9}
\end{figure}

In this subsection, we present simulation results of data utilities calculated by (\ref{utility}) and total costs calculated by (\ref{total_cost}), to analyze their impacts on the objective value. Fig. \ref{fig6}, Fig. \ref{fig7}, Fig. \ref{fig8}, and Fig. \ref{fig9} illustrate data utilities and energy costs of the proposed versatile algorithms under different values of $\phi_{u}$.
Thereof, the $x$-axis represents the number of edge servers, while the left $y$-axis depicts the results of the data utility function for dashed lines and the right $y$-axis demonstrates that of total cost values for solid lines.
It can be observed that the costs in these four figures are the same, while the range of data utility values decreases with the increasing $\phi_{u}$, resulting in a decrease in overall performance as shown in Fig. \ref{fig4}.
Specifically, except for Baseline 3, all strategies result in low data utility values when the number of edge servers is small. Then, with the increase in resources brought by the growing number of edge servers, the data utility values of these algorithms increase and reach their peak levels based on the system resources they can access due to the user digital twin association results. This peak level indicates that approximately each edge server serves only one user digital twin. Conversely, due to adopting a random allocation strategy, Baseline 3 maintains its data utility value at the same level across all scenarios characterized by the number of edge servers. However, this strategy results in much higher additional cost expenses, decreasing the overall performance, especially compared with Baseline 2, which adopts an identical association strategy.

Regarding the trend of data utilities and costs among Baseline 1, 2, 4, and our proposed method, it can be observed that our strategy and Baseline 1 can expense less costs for user digital twin association both when the number of edge servers is low and high. With the increase in the number of edge servers, these two algorithms incur slightly more costs but yield significantly more benefits in data utility, resulting in an overall performance increase. Due to our method's ability to better confine the association results within a reasonable range in each iteration, our method can converge more quickly, leading to a superior performance.

\subsection{Data Utility versus episode numbers}
\label{sec5d}

\begin{figure}
\centering
\subfigure[$\phi_{u} = 0.0$]{
\begin{minipage}[t]{0.45\linewidth}
    \centering
    \includegraphics[width=1\textwidth]{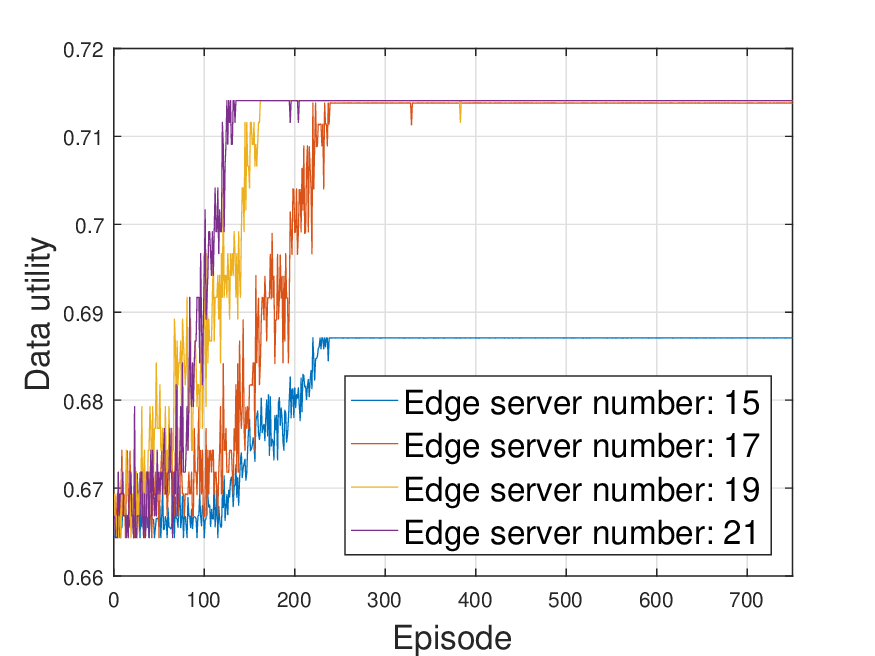}
\end{minipage}
}
\subfigure[$\phi_{u} = 0.2$]{
\begin{minipage}[t]{0.45\linewidth}
    \centering
    \includegraphics[width=1\textwidth]{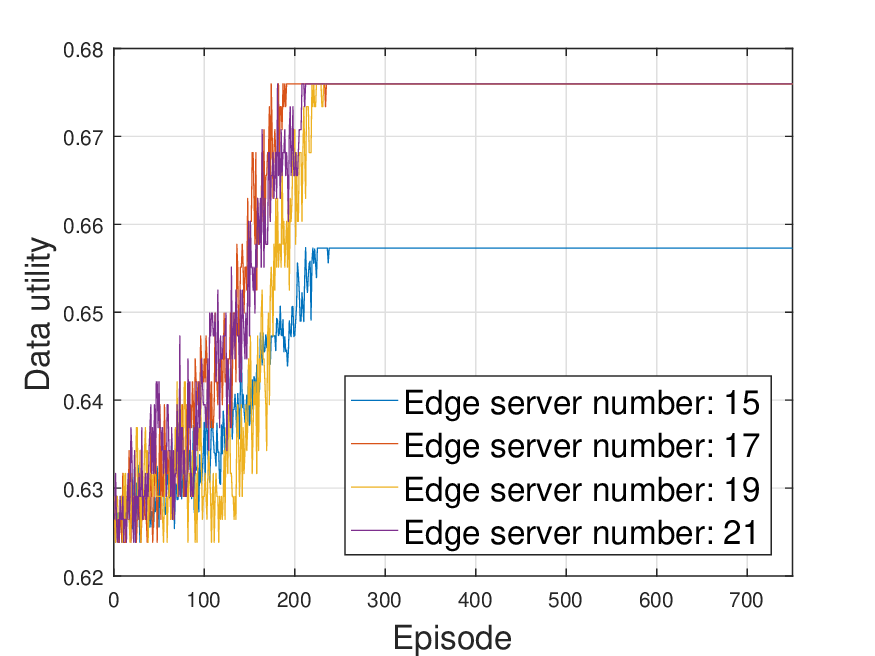}
\end{minipage}
}
\subfigure[$\phi_{u} = 0.4$]{
\begin{minipage}[t]{0.45\linewidth}
    \centering
    \includegraphics[width=1\textwidth]{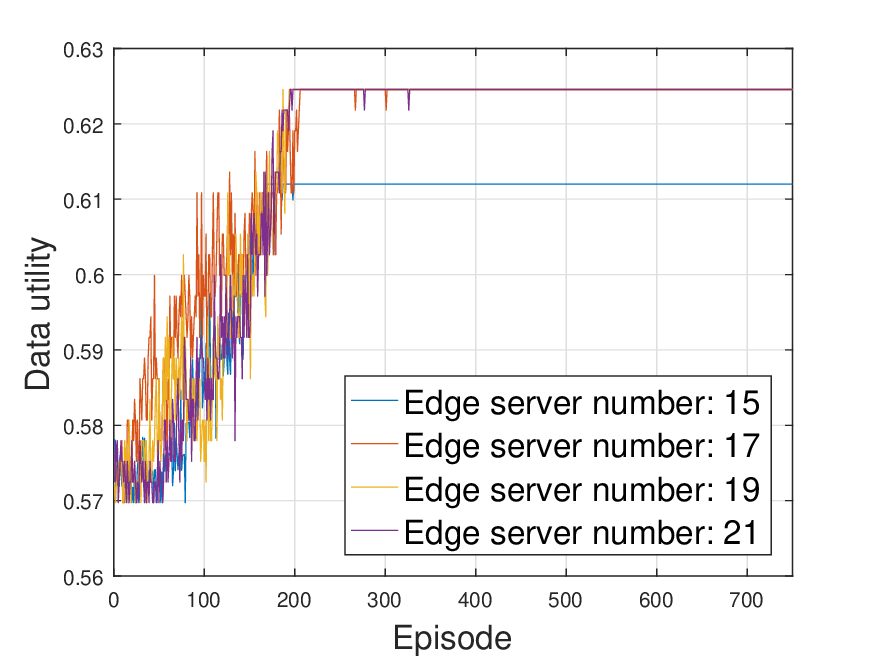}
\end{minipage}
}
\subfigure[$\phi_{u} = 0.6$]{
\begin{minipage}[t]{0.45\linewidth}
    \centering
    \includegraphics[width=1\textwidth]{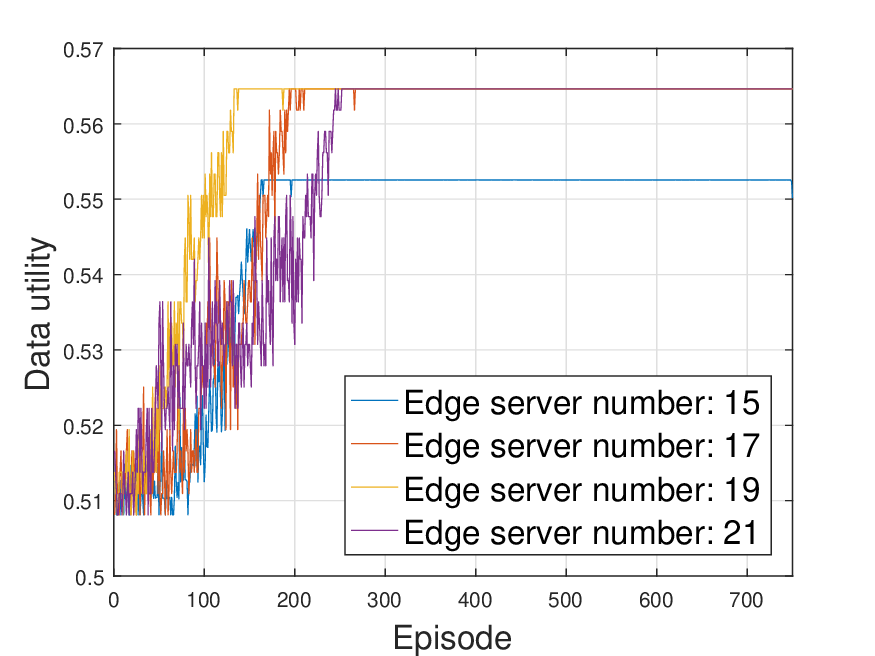}
\end{minipage}
}
\centering
\caption{Data utility of PPO algorithm with 20 users.}
\label{fig10}
\end{figure}

In this subsection, we evaluate the instances of data utility for our proposed algorithm along with the episodes under various parameter settings, as illustrated in Fig. \ref{fig10}. The parameter setting is the same as that in subsection \ref{sec5a}, except the $y$-axis representing the data utility in each episode calculated by (\ref{utility}), and the number of edge servers is set as $15$, $17$, $19$, and $21$ to avoid redundancy. From each of the sub-figures, we observe that the data utility of our proposed algorithm initially rises rapidly, then stabilizes at a specific value. Comparing with Fig. \ref{fig4}, it can be noted that the episode at which the data utility reaches its peak is the same as that of algorithm convergence, around $200$ episodes. When the number of edge servers becomes large enough, the data utility becomes the largest, representing the full allocation of historical data for improved performance given the system resources. Conversely, when the number of edge servers is comparatively small, such as $15$, the peak that the data utility value reached is relatively small, indicating the trade-off between the training performance and corresponding cost reduction.

Upon comparing the four sub-figures, it is evident that data utilities start at different values, depending on their EMD values $\phi_{u}$. This discrepancy arises because at the episode's onset, the performance of user digital twin association is poor, leading to limited available resources. Consequently, the corresponding historical data allocations are at a small value. However, the results brought by allocation strategies in different EMD value $\phi_{u}$ are varying. Specifically, even when the resources are limited, the allocation results when $\phi_{u}$ is small are larger than those when $\phi_{u}$ is large, leading to better performance in data utilities.

\section{Conclusion and Future Work}
\label{sec6}

In this paper,
we explored the long-term requirements of the DITEN and its conflicting concerns related to users' computing ability and the network's privacy leakage. To handle these issues, we investigated the FL procedure implemented on the DITEN and demonstrated a closed-form data utility function for predicting training accuracy in the long-term FL procedure. We termed this prediction the data utility.
The proof of the convergence for FL was also given. Based on this, we studied a maximization problem to strike a balance between the benefits of the FL on the DITEN and associated energy costs.
Specifically, the user digital twin association and the historical data allocation were optimized to maximize the data utility while minimizing associated energy costs.
To render the tractability of the formulated problem, we proposed an optimization-driven learning algorithm. We first decoupled the historical data allocation problem from the original problem
and proved this sub-problem to be convex. We then adopted an online DRL-based algorithm to address user digital twin association. Numerical results showed that our proposed algorithm can effectively find high-quality solutions that outperform the baseline approaches.

Note that this work provided a comprehensive perspective to integrate long-term DITEN maintenance and FL digital twin tasks. However, it also generalizes detailed digital twin mechanisms, such as the digital twin synchronizing procedure, which may impact its implementation. Therefore, in future work, we will delve into the specifics of digital twin synchronization. Specifically, by jointly considering the age of information and digital twin distortion, we will analyze the impact of synchronization details on the DITEN in terms of digital twin freshness and digital twin fidelity.


\begin{appendix}
\subsection{Proof of Theorem \ref{ConvergeTheorem}}
\label{ProofConverge}

From the training procedure, we know that
\begin{align}
&\omega^{t,i+1} \nonumber\\
& = \frac{1}{D_{t}^{\text{dat}}}\sum_{s\in\cS}D_{s,t}^{\text{dat}}\cdot \omega_{s}^{t,i+1} \nonumber\\
& = \frac{1}{D_{t}^{\text{dat}}}\sum_{s\in\cS}D_{s,t}^{\text{dat}}\cdot \left[\omega^{t,i}-\frac{\chi}{D_{s,t}^{\text{dat}}}\sum_{\{x,y\}\in\cD_{s,t}^{\text{dat}}}\nabla f(\omega^{t,i},x,y)\right]\nonumber\\
& = \omega^{t,i} - \frac{\chi}{D_{t}^{\text{dat}}}\sum_{s\in\cS}\sum_{\{x,y\}\in\cD_{s,t}^{\text{dat}}}\nabla f(\omega^{t,i},x,y)\nonumber\\
& = \omega^{t,i} - \frac{\chi}{D_{t}^{\text{dat}}}\sum_{\{x,y\}\in\cD_{t}^{\text{dat}}}\nabla f(\omega^{t,i},x,y)\nonumber\\
& = \omega^{t,i} - \chi\nabla F(\omega^{t,i}).
\end{align}
Therefore,
\begin{equation}
\omega^{t,i+1} - \omega^{t,i} = - \chi\nabla F(\omega^{t,i}).
\label{weightMinues}
\end{equation}
Then, we rewrite $F(\omega^{t,i})$ using the second-order Taylor expansion, which is depicted by
\begin{align}
&F(\omega^{t,i+1})\nonumber\\
& = F(\omega^{t,i}) + (\omega^{t,i+1} - \omega^{t,i})^{\text{T}}\nabla F(\omega^{t,i})\nonumber\\
&+\frac{1}{2}(\omega^{t,i+1} - \omega^{t,i})^{\text{T}}\nabla^{2} F(\omega^{t,i})(\omega^{t,i+1} - \omega^{t,i}).
\end{align}
Based on the third assumption, we can get
\begin{align}
&F(\omega^{t,i+1})\leq F(\omega^{t,i}) + (\omega^{t,i+1} - \omega^{t,i})^{\text{T}}\nabla F(\omega^{t,i})\nonumber\\
&+\frac{L}{2}||\omega^{t,i+1} - \omega^{t,i}||^{2}.
\label{convergeTaylor}
\end{align}
By substituting (\ref{weightMinues}) into (\ref{convergeTaylor}), we have
\begin{align}
&F(\omega^{t,i+1})\leq F(\omega^{t,i}) - \chi\nabla F(\omega^{t,i})^{\text{T}}\nabla F(\omega^{t,i})\nonumber\\
&+\frac{L\cdot\chi^{2}}{2}||\nabla F(\omega^{t,i})||^{2}\nonumber\\
&=F(\omega^{t,i}) - (\chi-\frac{L\cdot\chi^{2}}{2})||\nabla F(\omega^{t,i})||^{2}.
\end{align}
With the given $\omega^{t,*}$, we have
\begin{align}
&F(\omega^{t,i+1}) - F(\omega^{t,*}) \nonumber\\
&\leq F(\omega^{t,i}) - F(\omega^{t,*}) - (\chi-\frac{L\cdot\chi^{2}}{2})||\nabla F(\omega^{t,i})||^{2}.
\end{align}
Based on the second and third assumptions, there is a result \cite{ConvergenceResult}:
\begin{equation}
||\nabla F(\omega^{t,i})||^{2}\geq2\psi(F(\omega^{t,i}) - F(\omega^{t,*})).
\end{equation}
Consequently,
\begin{align}
&F(\omega^{t,i+1}) - F(\omega^{t,*}) \nonumber\\
&\leq F(\omega^{t,i}) - F(\omega^{t,*})\nonumber\\
&-[\psi\chi(2-L\cdot\chi)](F(\omega^{t,i}) - F(\omega^{t,*}))\nonumber\\
&\leq (1-2\psi\chi+L\psi\chi^{2})^{i+1}(F(\omega^{t,0}) - F(\omega^{t,*})).
\end{align}
Due to the fact that $1-x\leq e^{-x}$, we have
\begin{align}
&F(\omega^{t,i+1}) - F(\omega^{t,*})\leq \nonumber\\
&e^{-(i+1)(2\psi\chi-L\psi\chi^{2})}(F(\omega^{t,0}) - F(\omega^{t,*})).
\end{align}
For the fine-tuning, the proof is identical. The theorem \ref{ConvergeTheorem} is proved.
\end{appendix}

\bibliographystyle{IEEEtran}

\bibliography{reference}

\end{document}